\documentclass{aastex62}

\usepackage{amsmath}

\received{}
\revised{}
\accepted{}

\submitjournal{PASP}

\shorttitle{Contact Binary}
\shortauthors{Junhui Liu et al.}

\begin{document}

\title{An optical and X-ray study of the contact binary, BH Cassiopeiae}

\correspondingauthor{Ali Esamdin, Yu Zhang}
\email{aliyi@xao.ac.cn, zhy@xao.ac.cn}

\author[0000-0002-7600-1670]{Junhui Liu}
\affil{Xinjiang Astronomical Observatory, Chinese Academy of Sciences, Urumqi 830011,
Xinjiang, China}
\affiliation{University of Chinese Academy of Sciences, Beijing 100049, Beijing, China}

\author{Ali Esamdin}
\affiliation{Xinjiang Astronomical Observatory, Chinese Academy of Sciences, Urumqi 830011,
Xinjiang, China}
\affiliation{University of Chinese Academy of Sciences, Beijing 100049, Beijing, China}

\author{Yu Zhang}
\affiliation{Xinjiang Astronomical Observatory, Chinese Academy of Sciences, Urumqi 830011,
Xinjiang, China}

\author{Chin-Ping Hu}
\altaffiliation{JSPS International Research Fellow}
\affiliation{Department of Astronomy, Kyoto University, Oiwake-cho, Sakyo-ku, Kyoto 606-8502, Japan}

\author{Tingting Chen}
\affiliation{Xinjiang Astronomical Observatory, Chinese Academy of Sciences, Urumqi 830011,
Xinjiang, China}
\affiliation{University of Chinese Academy of Sciences, Beijing 100049, Beijing, China}

\author{Junbo Zhang}
\affiliation{Key Laboratory of Optical Astronomy, National Astronomical Observatories, Chinese Academy of Sciences, A20 Datun Road,
Chaoyang District, Beijing 100012, China}
\affiliation{University of Chinese Academy of Sciences, Beijing 100049, Beijing, China}

\author{Jinzhong Liu}
\affiliation{Xinjiang Astronomical Observatory, Chinese Academy of Sciences, Urumqi 830011,
Xinjiang, China}

\author{Zixi Li}
\affiliation{College of Physics and Electronic Engineering, Xinjiang Normal University, Urumqi, Xinjiang, 830054, China}
\affiliation{Xinjiang Astronomical Observatory, Chinese Academy of Sciences, Urumqi 830011,
Xinjiang, China}

\author{Juanjuan Ren}
\affiliation{Key Laboratory of Optical Astronomy, National Astronomical Observatories, Chinese Academy of Sciences, A20 Datun Road,
Chaoyang District, Beijing 100012, China}

\author{Jie Zheng}
\affiliation{Key Laboratory of Optical Astronomy, National Astronomical Observatories, Chinese Academy of Sciences, A20 Datun Road,
Chaoyang District, Beijing 100012, China}

\author{Hubiao Niu}
\affiliation{Department of Astronomy, Beijing Normal University, Beijing 100875,
China}
\affiliation{Xinjiang Astronomical Observatory, Chinese Academy of Sciences, Urumqi 830011,
Xinjiang, China}

\author{Chunhai Bai}
\affiliation{Xinjiang Astronomical Observatory, Chinese Academy of Sciences, Urumqi 830011,
Xinjiang, China}

\author{Liang Ge}
\affiliation{Xinjiang Astronomical Observatory, Chinese Academy of Sciences, Urumqi 830011,
Xinjiang, China}

\begin{abstract}


We present the revised and high quality CCD time series observations in 2017 and 2018 of the W UMa-type contact binary BH~Cas.
The combination of our new determinations of minimum times and those collected from the literatures
reveal a long-term period increase with a rate of $\textit{dP/dt}\,=\,+3.27 \times 10^{-7} \textrm{day}\,\textrm{yr}^{-1}$,
which can be accounted for by angular momentum transfer as the less-massive component loses its mass to the more massive one.
One cyclical oscillation ($A'=0.00300$~day and $P'=20.09$~yr) superimposed on the long-term period increase tendency is attributed to
one additional object that orbits BH Cas. The low-resolution spectra of this source are observed at phase $\sim$0.0 and $\sim$0.5 in 2018 and 2019,
and the spectral types are identified as K3V$\,\pm\,$1 and G8V$\,\pm\,$2. According to the luminosity contributions and spectra at these phases, the spectral type of the primary star is close to K3V$\,\pm\,$1. We infer a cool spot on the hotter and less-massive component to fit the asymmetry of light curves. In addition, changes in the location, temperature, and area of spots on the secondary star in different observations may indicate that the magnetic field activity of the secondary is more active than that of the primary. Such strong spot activities and optical flare are both detected at the same
time from a W UMa-type star for the first time. With the \emph{XMM-Newton}
observations taken in 2003 and 2012, the X-ray light curve is obtained and the X-ray spectra can
be well described with a black-body or a bremsstrahlung model. The soft X-ray luminosity of BH Cas is estimated as 8.8\,--\,9.7$\,\times\,10^{30}$\,erg\,s$^{-1}$.

\end{abstract}


\section{Introduction} \label{sec:intro}

Close binaries are divided into Algol (EA), $\beta$ Lyrae (EB) and W UMa (EW) types
\citep{1998stel.conf...81M} according to the shape of their light curves.
W UMa-type binaries, which have been studied for more than one century \citep{1903PA.....11..138M},
are one of the EW-type with a relatively high frequency of occurrence \citep{1948HarMo...7..249S,2008MNRAS.385.2239N}.
Many phenomena, however, remain unexplained.  For example, the O'Connell effect
\citep{1951PRCO....2...85O, 1968AJ.....73..708M} shows two maxima times of the light curves with
different luminosities. Based on the temperatures and masses, the W UMa-type can be further
divided into the W-subtype and the A-subtype \citep{1970VA.....12..217B}. For the W-subtype, the primary
minima of the light curves are caused by the more massive component transiting the less massive hotter one, whereas
for the A-subtype, the situation is reverse.
These two subtypes of W UMa-type are generally considered to have an evolutionary relationship with each other,
but the direction of evolution between them remains controversial.
For example, \citet{2004MNRAS.351..137L}, based on the energy transfer in binary systems, infers that the A-subtype
is the later evolution stage of the W-subtype. \citet{2006MNRAS.370L..29G}, on the other hand, propose
the opposite evolutionary sequence by utilizing statistics of the mass and angular momentum of binary systems.

Many W UMa-type binaries are X-ray emitters. Some of their fundamental features, including high chromospheric and
coronal activity \citep{ 2006ApJ...650.1119H, 2016AJ....151..170H, 2015MNRAS.446..510K}, and
synchronous fast-rotation common envelopes \citep{2004A&A...415.1113G, 2006AJ....131..990C}, are
believed to be the origin of the X-ray emission. Their X-ray intensities are related to
their orbital periods and spectral types \citep{2001A&A...370..157S, 2006AJ....131..990C}.
The studies of 2MASS J11201034-2201340 \citep{2016AJ....151..170H} and VW Cep \citep{2006ApJ...650.1119H}
indicate that the X-ray light curves do not show obvious occultation or modulation as optical light curves do.
The investigation of VW Cep and YY Eri \citep{2007IAUS..240..719V} suggests that in a UMa binary system,
the massive component dominates the magnetic activity.

BH Cassiopeiae (RA=00$^{h}$21$^{m}$21.4$^{s}$, Dec=+59$^{\circ}$09$^{\prime}$05$^{\prime\prime}$.2, J2000) was
recognized by \citet{1931AN....243..115B} as a variable star. \citet{KukarkinB1938} classified it as a W UMa-type
variable star with a period of $\sim0.5$~days and an amplitude of $\sim0.4$~mag.
\citet{1999AJ....117.2503M} presented the light curves in \textit{U}, \textit{B} and \textit{V} bands and
the radial velocity curves. He obtained basic parameters (e.g., the inclination, temperatures, and fractional luminosity)
of BH Cas and further classified it as a W-subtype W UMa system. However, he corrected for data taken with different
telescopes by assuming all \textit{U} band maxima to be approximately equal, thereby preventing him from detection of
the O'Connell effect \citep{2001A&A...374..164Z}. \citet{2001A&A...374..164Z} carried out additional photometric
observations in \textit{R} and \textit{I} bands, and found no O'Connell effect.
\citet{2001hell.confE..77N} also detected no significant O'Connell effect in their light curves of BH Cas.
For the temperature of BH Cas, \citet{1999AJ....117.2503M} derived an effective temperature about $4600\pm400$~K
by the color index, and later concluded the temperature of the secondary star to be $4980\pm100$~K and
the temperature of primary star to be $4790\pm100$~K. \citet{2001A&A...374..164Z} set the temperature of the
secondary as 6000~K based on a classification spectrum F8$\pm2$ for BH Cas, and derived the temperature of the primary as
$5550\pm22$~K.

Information regarding the orbital period change is vital in derivation of the physical properties
of contact binaries.  By conducting the Observed minus Calculated (O$-$C) analysis for BH Cas,
\citet{2001A&A...374..164Z} noted that the parameters of the linear fit for O$-$C analysis
(data covering from 1994 to 2001)
were positive, and those of the parabolic fit were close to zero.
However, \citet{2001MNRAS.328..635Q} showed the O$-$C to have a significant upward trend
with a rate of $dP/dt = +1.17 \times10^{-6}$~d~yr$^{-1}$ based on the data obtained
from 1994 to 1999, and notwithstanding the relatively short time interval of the observations,
commented the rate to be unusually large for a binary system.
Using data taken from 1994 to 2008, \citet{2009Obs...129...88A} measured the linear elements to be nearly 0.

In this paper, we report the revised high-quality CCD photometric data of BH\,Cas, which when supplemented
with minimum times in the light curves collected from the literature, led to an improved determination of
the physical parameters of the system.  New spectral typing provides more reliable estimation on the
stellar temperatures than before, whereas the X-ray light curves and spectra set constraints
on the emission mechanisms.  We present the optical photometric, spectral and X-ray observations of BH\,Cas
in Section \ref{sec:processes}. The O$-$C analysis, optical light curve fitting, and X-ray spectral
fitting are given in Section \ref{sec:analysis}. We discuss our results in Section \ref{sec:Discu}
and give a summary of our work in Section \ref{sec:Summary}.

\section{Observations, Data Reductions and Results} \label{sec:processes}

\subsection{Optical Observation and Data Reduction}\label{sec:optical}
\subsubsection{Photometry}\label{sec:Photometry}

BH Cas was observed using the Nanshan 1-meter telescope \citep{2016RAA....16..154S,2018Ap&SS.363...68M} of
Xinjiang Astronomical Observatory from Oct. 20, 2017 to Dec. 29, 2018. The telescope was equipped with an
E2V CCD203-82 (blue) chip CCD camera, which has $4096 \times 4136$ pixels, with a pixel size of
1.125 arcsec. A square area of $1200 \times 1200$ pixels near the center of the CCD chip was used,
corresponding to a field of view of 22$^{\prime}$.5\,$\times$\,22$^{\prime}$.5. A set of Johnson-Cousins
\textit{B}, \textit{V}, \textit{R} and \textit{I} filters was used, with typical exposure times of
20~s for \textit{B}, 15~s for \textit{V}, 8~s for \textit{R} and 17~s for \textit{I}.  For all the
observations, the same gain, bin values and readout mode were used.

\begin{figure}
\centering
\includegraphics[width = 6cm]{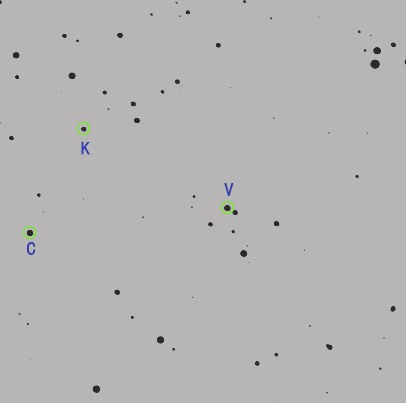}
\caption{ 5.1$^{\prime}$\,$\times$\,5.1$^{\prime}$ image in \textit{B} band. The target BH Cas, the
	comparison star and the check star are marked, respectively, with ``V'', ``C'' and ``K''. }
\label{fig:positionofstar}
\end{figure}

The aperture photometry package PHOT of IRAF\footnote{IRAF is distributed by the National Optical Astronomy
Observatory, which is operated by the Association of Universities for Research in Astronomy, Inc.,
under cooperative agreement with National Science Foundation} is utilized to reduce the CCD data.
Figure \ref{fig:positionofstar} demonstrates a part of the observed frame, in which the target,
comparison and check stars are labeled.  The magnitudes and positions of the comparison and check stars
are close to BH Cas. The color of the comparison star is almost the same as that of BH \,Cas.
The coordinates and visual magnitudes of these stars are listed in table \ref{table:Coordinates}.
For differential magnitudes, the IRAF command \textit{pdump} is used to derive the instrumental magnitudes
and their errors, and \textit{helect} is used to extract the Heliocentric Julian Date (HJD) of each frame.
Then, the new times of the primary (type I) and secondary (type II) minima in different bands,
listed in Table \ref{table:new_mini}, are derived by parabolic least-squares fitting to the light curves.
 We started out with the equation
$\textmd{Min.I}$\,=\,($\textmd{HJD$_{0}$}$)\,2449998.6182(3)\,+\,0.$^{d}$40589171(5)\,$\times$\,\textit{E} as
per given in \citet{2009Obs...129...88A} to obtain the orbital phases, but there is a slight shift in phase.
Therefore, we chose one of the minimum times we derived as the new HJD$_{0}$.
The period is preserved, and the revised linear ephemeris becomes,
\begin{equation}\label{equ:myminI}
    \textmd{Min.I}=(\textmd{HJD$_{0}$})\,2458124.1950(1)\,+\,0.^{d}40589171(5)\,\times\,E.
\end{equation}

\begin{figure}
\centering

\includegraphics[width = 14cm]{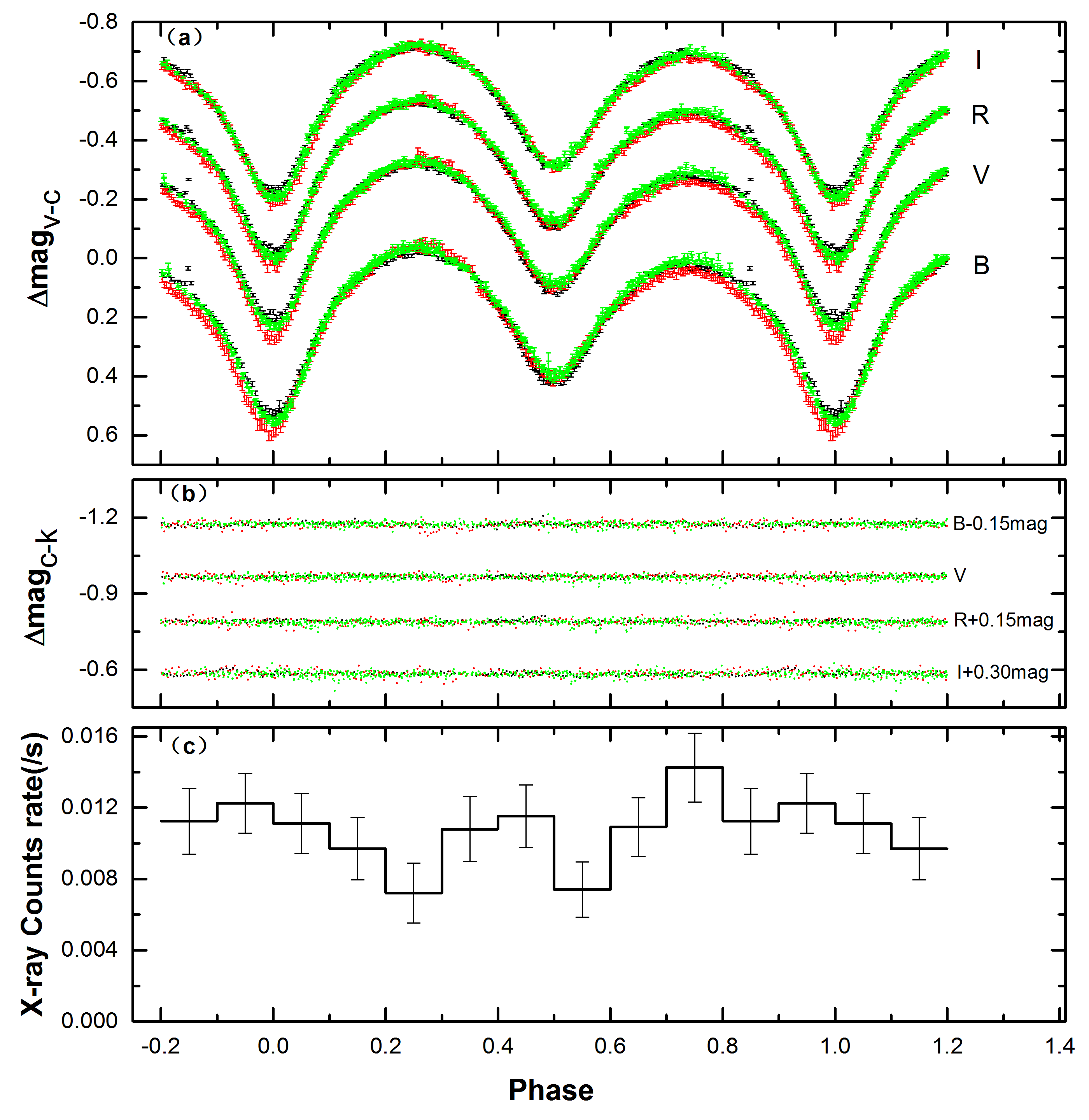}
\caption{(a): \textit{BVRI} bands phase-folded light curves for BH Cas. The error bars are set as 2$\sigma$ instrumental magnitudes uncertainties. A clear flare was detected in Oct. 21, 2017. (b): The magnitude differences of the comparison (\textit{C}) and the check star (\textit{K}). These colors represent the same meaning as panel (a). (c): The phase-folded X-ray light curve obtained by \emph{XMM-Newton} in the 0.2--12keV energy range (Section \ref{sec:X-ray_observation}). The error bars are 1$\sigma$ statistical uncertainties.}
\label{fig:LC}
\end{figure}

Figure \ref{fig:LC}(a) exhibits the phase-folded optical light curves of BH Cas observed on Oct. 20, 21(black),
Dec. 21(red), 2017 and Jan. 4, 5, 6, 7(green), 2018. In each band of the light curves, the data obtained in these three observation seasons deviate more from each other in the phase range from 0.6 to 1.06 compared with that in other phase ranges. We define the primary (MaxI) and secondary (MaxII) maximum as that following the primary and secondary minima, respectively. The MaxI and MaxII magnitudes, both obtained from the same nights by parabolic least-squares fitting to the light curves, are listed in Table~\ref{table:maximum_magnitude}. It can been seen that the magnitude difference between the amplitude of the primary and the secondary maxima (MaxI-MaxII) change over observations. Note a clear flare event detected on Oct. 21, 2017. Figure \ref{fig:LC}(b), showing the differential magnitudes between the comparison and the check stars,
manifests that the observed variations in the optical light curves of BH\,Cas are actual and trustworthy.

 \begin{table*}
  \caption{Coordinates and visual magnitudes of the target, comparison and check stars}
   \begin{center}
   \begin{tabular}{ccccc}\hline \hline

Stars                              &$\alpha_{2000}$            & $\delta_{2000}$                                  &  $V$      &$B-V$        \\
                                   &                           &                                                  & (mag)     &($\Delta$mag) \\  \hline
BH Cas(V)                          &00$^{h}$21$^{m}$21.4$^{s}$ & +59$^{\circ}$09$^{\prime}$05$^{\prime\prime}$.2  &12.77$\pm$0.22  &       0.49 \\
2MASS J00211690+5912483(C)         &00$^{h}$21$^{m}$16.9$^{s}$ & +59$^{\circ}$12$^{\prime}$48$^{\prime\prime}$.3  &12.92$\pm$0.06  &       0.50 \\
2MASS J00213244+5911504(K)         &00$^{h}$21$^{m}$32.4$^{s}$ & +59$^{\circ}$11$^{\prime}$50$^{\prime\prime}$.4  &13.90$\pm$0.03  &       0.58 \\
\hline\noalign{\smallskip}
  \end{tabular}
  \end{center}
  \label{table:Coordinates}
\end{table*}

\begin{table*}
  \caption{New minimum times of BH Cas observed in different bands through this work. }
   \begin{center}
   \begin{tabular}{cccccccccc}\hline \hline
Date&HJD &Type &Error  & band & Date&HJD &Type &Error  & band \\
    & 2400000+ &       &       &    & &2400000+  &      &      \\
\hline
2017.10.20&58047.27975&II&0.00024&\textit{B}        &2018.02.07& 58157.07476 & I  & 0.00050 & \textit{B} \\
 & 58047.27954 & II  & 0.00012 & \textit{V}         &       & 58157.07417 & I  & 0.00031 & \textit{V} \\
 & 58047.27931 & II  & 0.00037 & \textit{R}         &       & 58157.07414 & I  & 0.00032 & \textit{R} \\
 & 58047.27912 & II  & 0.00035 & \textit{I}         &2018.02.08& 58158.08912 & II  & 0.00020 & \textit{B} \\
2017.10.21& 58048.29337 & I  & 0.00025 & \textit{V} &       & 58158.08880 & II  & 0.00017 & \textit{V} \\
 & 58048.29293 & I  & 0.00034 & \textit{R}          &       & 58158.08922 & II  & 0.00026 & \textit{R} \\
 & 58048.29332 & I  & 0.00051 & \textit{I}          &       & 58158.08891 & II  & 0.00029 & \textit{I} \\
2017.12.01& 58089.08652 & II  & 0.00044& \textit{B} &2018.11.03& 58426.18572 & I  & 0.00036 & \textit{B} \\
 & 58089.08612 & II  & 0.00030 & \textit{V}         &       & 58426.18565 & I  & 0.00022 & \textit{V} \\
 & 58089.08594 & II  & 0.00089 & \textit{R}         &       & 58426.18406 & I  & 0.00020 & \textit{R} \\
 & 58089.08641 & II  & 0.00055 & \textit{I}         &       & 58426.18561 & I  & 0.00081 & \textit{I} \\
2017.12.01& 58089.28710 & I  & 0.00044& \textit{B}  &2018.11.03& 58426.38775 & II  & 0.00044 & \textit{B} \\
 & 58089.28772 & I  & 0.00067 & \textit{V}          &       & 58426.38691 & II  & 0.00035 & \textit{V} \\
 & 58089.28732 & I  & 0.00034 & \textit{R}          &       & 58426.38804 & II  & 0.00021 & \textit{R} \\
 & 58089.28854 & I  & 0.00079 & \textit{I}          &       & 58426.38791 & II  & 0.00077 & \textit{I} \\
2018.01.04& 58123.18243 &II & 0.00032&\textit{B}    &2018.11.04& 58427.20096 & I  & 0.00078 & \textit{B} \\
 & 58123.18230 & II  & 0.00022 & \textit{V}         &  & 58427.20033 & I  & 0.00049 & \textit{V} \\
 & 58123.18175 & II  & 0.00038 & \textit{R}         &        & 58427.20074 & I  & 0.00022 & \textit{R} \\
 & 58123.18128 & II  & 0.00020 & \textit{I}         & & 58427.20093 & I  & 0.00093 & \textit{I} \\
2018.01.04& 58123.99454 & II  & 0.00032 &\textit{V} &2018.11.04& 58427.40276 & I  & 0.00031 & \textit{B} \\
2018.01.05& 58124.19429 & I  & 0.00023&\textit{B}   & & 58427.40256 & I  & 0.00013 & \textit{V} \\
 & 58124.19474 & I  & 0.00016 & \textit{V}          & & 58427.40276 & I  & 0.00073 & \textit{R} \\
 & 58124.19505 & I  & 0.00026 & \textit{R}          & & 58427.40318 & I  & 0.00019 & \textit{I} \\
 & 58124.19576 & I  & 0.00022 & \textit{I}          &2018.11.29& 58452.16318 & I  & 0.00094 & \textit{B} \\
2018.01.06& 58125.00740 & I  & 0.00075 & \textit{B} & & 58452.16351 & I  & 0.00023 & \textit{V} \\
 & 58125.00714 & I  & 0.00010 & \textit{V}          & & 58452.16339 & I  & 0.00019 & \textit{R} \\
 & 58125.00753 & I  & 0.00086 & \textit{R}          & & 58452.16373 & I  & 0.00051 & \textit{I} \\
 & 58125.00835 & I  & 0.00054 & \textit{I}          &2018.12.29& 58482.19930  &I  & 0.00086 & \textit{B} \\
2018.01.06& 58125.21119 & II  & 0.00030&\textit{B}  & & 58482.19940 & I  & 0.00068 & \textit{V} \\
 & 58125.21102 & II  & 0.00020 & \textit{V}         & & 58482.19951 & I  & 0.00082 & \textit{R} \\
 & 58125.20963 & II  & 0.00022 & \textit{R}         & & 58482.19960 & I  & 0.00053 & \textit{I} \\
 & 58125.21017 & II  & 0.00021 & \textit{I}    \\
2018.01.07&58126.22427&I&0.00020&\textit{B}    \\
 & 58126.22488 & I  & 0.00040 & \textit{V}    \\
 & 58126.22421 & I  & 0.00080 & \textit{R}   \\
 & 58126.22455 & I  & 0.00014 & \textit{I} \\

\hline\noalign{\smallskip}
  \end{tabular}
  \end{center}
\label{table:new_mini}
\end{table*}

\begin{table*}
  \caption{ The maximum magnitude values and the magnitude difference between the amplitude of the primary maxima and the following secondary of BH Cas in different bands.}
   \begin{center}
   \begin{tabular}{ccccccccl}\hline \hline
Date&MaxI&Error&MaxII&Error&MaxI-MaxII&Error& band  \\
     &($\Delta$mag) &                    & ($\Delta$mag) &            &    ($\Delta$mag)   &       \\ \hline
2017.12.01 & -0.042 & 0.008  &  0.039 & 0.007 &0.081  &0.015  & \textit{B}  \\
2017.12.01 & -0.327 & 0.008  & -0.259 & 0.009 &0.068  &0.017  & \textit{V}  \\
2017.12.01 & -0.538 & 0.009  & -0.480 & 0.006 &0.058  &0.015  & \textit{R}  \\
2017.12.01 & -0.721 & 0.010  & -0.682 & 0.006 &0.039  &0.016  & \textit{I}  \\
2018.01.04 & -0.034 & 0.005  & -0.001 & 0.010 &0.033  &0.015  & \textit{B}  \\
2018.01.04 & -0.333 & 0.005  & -0.301 & 0.008 &0.032  &0.013  & \textit{V}  \\
2018.01.04 & -0.534 & 0.005  & -0.503 & 0.007 &0.031  &0.013  & \textit{R}  \\
2018.01.04 & -0.718 & 0.004  & -0.701 & 0.008 &0.017  &0.013  & \textit{I}  \\
2018.01.05 & -0.047 & 0.008  &  0.011 & 0.006 &0.058  &0.014  & \textit{B}  \\
2018.01.05 & -0.325 & 0.007  & -0.289 & 0.004 &0.036  &0.012  & \textit{V}  \\
2018.01.05 & -0.534 & 0.010  & -0.498 & 0.003 &0.036  &0.013  & \textit{R}  \\
2018.01.05 & -0.724 & 0.006  & -0.690 & 0.004 &0.034  &0.010  & \textit{I}  \\

\hline\noalign{\smallskip}
  \end{tabular}
  \end{center}
  \label{table:maximum_magnitude}
\end{table*}

\subsubsection{Spectroscopy}\label{sec:Spectra}

\begin{figure}
  \centering
  \includegraphics[width = 14cm]{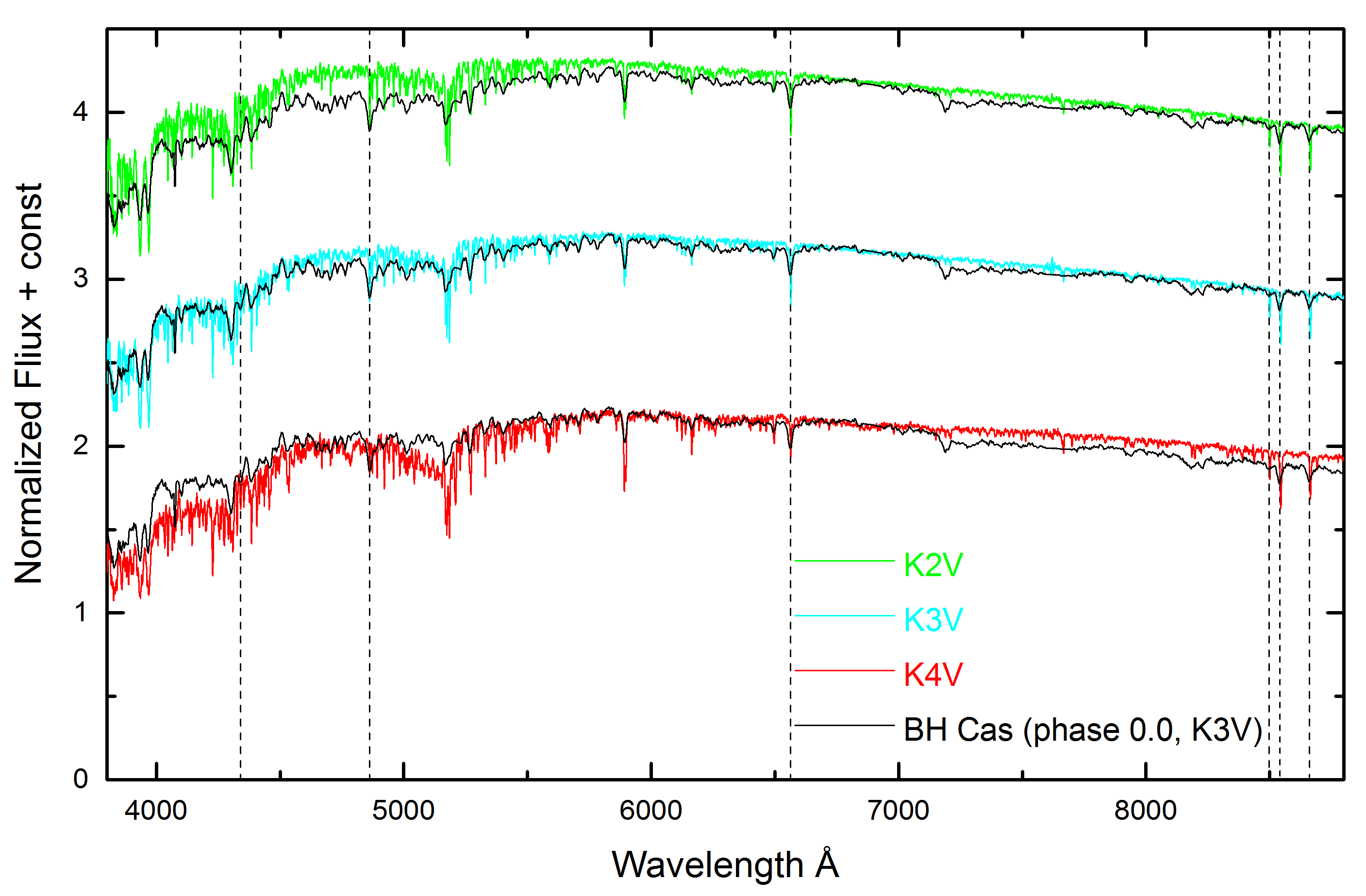}\\
  \caption{Spectrum of BH Cas at phase 0.0 compared with reference spectra of K2V--K4V. Best fit is obtained for K3V star with [Fe/H]$\,\sim\,-$1.0. All spectra are normalized at $\lambda$=6900{\AA}. From left to right, the black dashed lines represent H$\gamma$ (4341.6803{\AA}), H$\beta$ (4862.6778{\AA}), H$\alpha$ (6564.6127{\AA}) and triple lines (8498{\AA}, 8542{\AA} and 8662{\AA}) of CaII.}\label{fig:spectra_matchK3V}
\end{figure}
\begin{figure}
  \centering
  \includegraphics[width = 14cm]{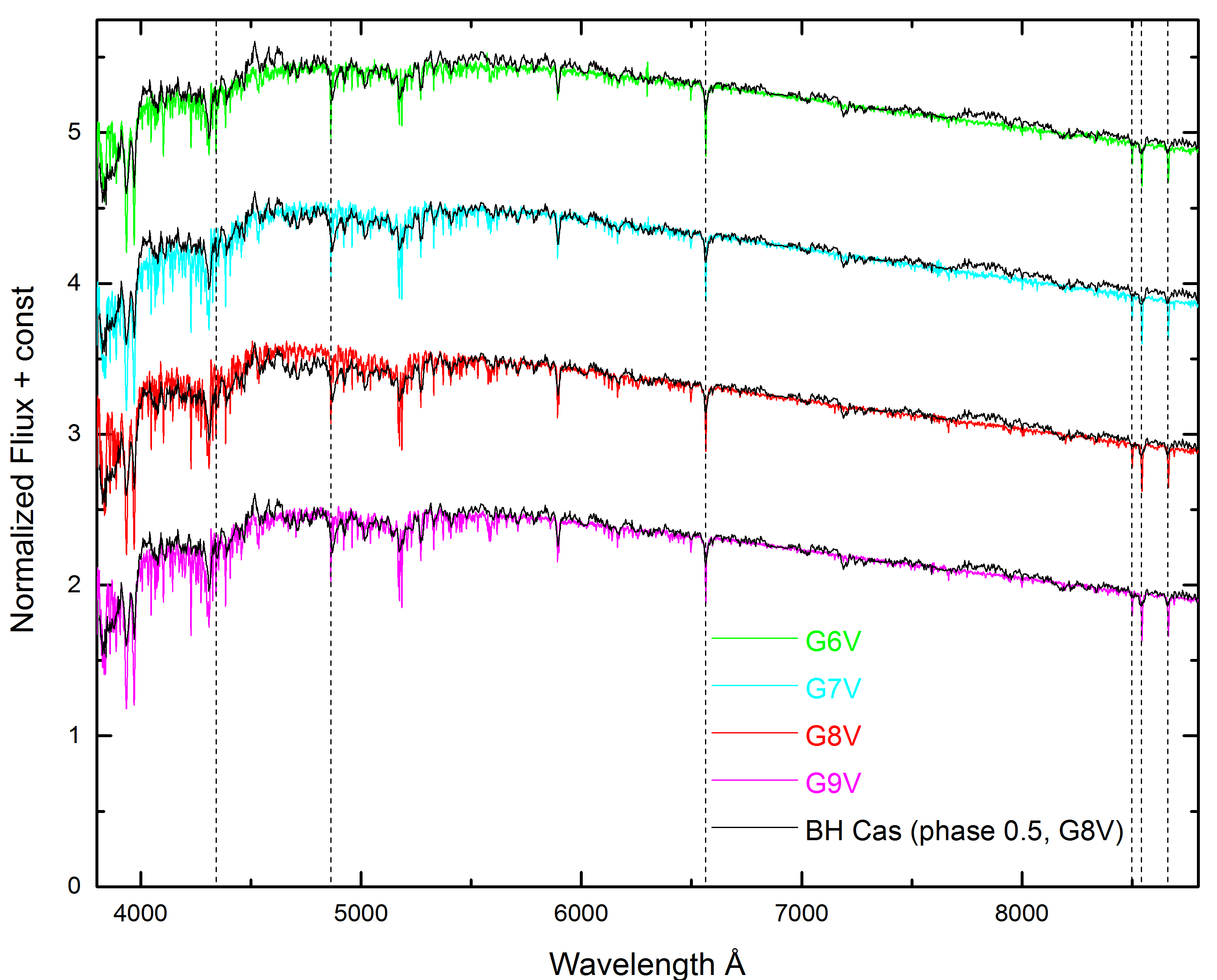}\\
  \caption{Spectrum of BH Cas at phase 0.5 compared with reference spectra of G6V--G9V. Best fit is obtained for G8V star with [Fe/H]$\,\sim\,-$1.0. All spectra are normalized at $\lambda$=6900{\AA}. From left to right, the black dashed lines represent H$\gamma$ (4341.6803{\AA}), H$\beta$ (4862.6778{\AA}), H$\alpha$ (6564.6127{\AA}) and triple lines (8498{\AA}, 8542{\AA} and 8662{\AA}) of CaII.}\label{fig:spectra_matchG8V}
\end{figure}

The low-resolution spectra of BH Cas were observed on November 22, 2018 and January 23, 2019 with the 2.16-meter telescope \citep{2016PASP..128k5005F} in the Xinglong Observatory, National Astronomical Observatories of China. The Beijing Faint Object Spectrograph and Camera, equipped with a 2048\,$\times$\,2048 pixels E2V CCD42-40 NIMO CCD, was used in these observations. A slit of 1.8$''$ and grating G4 was used to provide a wavelength coverage from 3800~\AA to 8800~\AA,
a linear dispersion of 198~\AA~mm$^{-1}$, and a spectral resolution of 4.45~\AA~pixel$^{-1}$.
The total exposure time on the target was 800~s. IRAF is used for data reduction following
standard spectral processing. Finally, the one-dimension spectra of BH Cas and the standard star
Hiltner~102 \citep{1974ApJ...193..135S} were extracted for flux calibration.
In Figure \ref{fig:spectra_matchK3V} and \ref{fig:spectra_matchG8V}, the spectra presented by
solid black lines are observed at phases $\sim$0.0 (HJD 2458445.067365) and $\sim$0.5 (HJD 2458506.948262),
respectively. The spectra were compared with SDSS/BOSS reference spectra of different spectral types
from \citet{2007AJ....134.2398C}. The PyHammer\footnote{https://github.com/BU- hammerTeam/PyHammer}
code \citep{2017ApJS..230...16K} was used to make the comparison, leading to the best fit of K3V$\pm$1 ($\sim$4850$\pm$150~K) and G8V$\pm$2 \citep[$\sim$5300$\pm$300~K]{2000asqu.book.....C,1988BAICz..39..329H}, both with [Fe/H]$\sim-1.0$ at phases 0.0 and 0.5, respectively.
 The positions of absorption lines H$\alpha$ (6564.6127{\AA}), H$\beta$ (4862.6778{\AA}), H$\gamma$ (4341.6803{\AA}) and triple lines (8498{\AA}, 8542{\AA} and 8662{\AA}) of CaII on the spectrum are represented by dashed lines.

According to geometric configuration \citep{2001A&A...374..164Z,1999AJ....117.2503M} of BH Cas, if we assume that two components are black bodies, the spectral temperature at phase 0.0 is the closest temperature to the primary through the whole phase range while that at phase 0.5 is the closest to the secondary. And since the light curves of BH Cas can be well fitted for a wide range of temperatures \citep{2001A&A...374..164Z}, the reliable initial values are important for the solution of light curve fitting. Therefore, it is reasonable to set the temperatures 4850~K and 5300~K derived by the spectrum of BH Cas at phases 0.0 and 0.5 as the initial temperatures of the primary and secondary in optical light curve fitting in Section \ref{sec:light curve model fitting}, respectively.

\subsection{X-ray Observations and Data Reduction}\label{sec:X-ray_observation}

The X-ray detection of BH Cas was firstly presented by \citet{1997MNRAS.291..709B} as a part of the X-ray observations
of the Local Group galaxy IC\,10. Later, \emph{XMM-Newton} \citep{2001A&A...365L...1J} observed BH Cas
with an $\sim40$~ks exposure on July 3, 2003 with ObsID: 0152260101 and $\sim130$~ks exposure on
August 18, 2012 with ObsID: 0693390101 \citep{2013ApJ...771L..44P}. We downloaded the data from XMM-Newton
Science Archive\footnote{http://nxsa.esac.esa.int/nxsa-web/}(XSA).
All the European Photon Imaging Cameras (EPICs), including two MOS cameras and PN CCD, were operated in
the full-frame mode with thin filters in these two observations. Details are listed in
Table \ref{table:XMM-Newton_observation_log}. In the following, we denote the X-ray observations in
2003 and 2012 as O$_{03}$ and O$_{12}$, respectively (see Table \ref{table:XMM-Newton_observation_log}).

We use XSA online Interacting Data Analysis\footnote{http://nxsa.esac.esa.int/nxsa-web/\#search} (IDA)
to perform pipeline data reduction. IDA processes the observation data files (ODF) with Science Analysis System (SAS)
and current calibration files on the Remote Interface for Science Analysis (RISA) server. The system
automatically generates filtered events lists, clean spectra and light curves. We extract source photons
from a circular region centered on BH Cas with a radius of 30$\arcsec$ to encircle ~80\% of source energy.
BH Cas lies on the boundary of two chips of the PN camera in O$_{12}$, so we do not use it. The background
events are extracted from a source-free 80$\arcsec$ circle near the source. In order to eliminate the
difference in results that may be caused by background selection, we also select the source-free
backgrounds of other sizes and positions, and find almost no difference in background events after
normalization. We do not take the pile-up effect into consideration, because the counts rate of
BH Cas $\sim0.01$~counts~s$^{-1}$ is far less than the minimum of EPICs needed for consideration.
For the event pattern and flag, the options are set to be "$\textrm{Pattern}\,<=\,12$"
and "Flag\,=\,0" for MOS, respectively, and "$\textrm{Pattern}\,<=\,4$", "Flag\,=\,0" for PN.
The energy range is selected to be between 0.2~keV to 12~keV. For the product type, firstly,
we set the option to be "Spectra". The files of source spectra, background spectra, spectral
response matrices for O$_{03}$ and O$_{12}$ are automatically generated by IDA. Then, the clean
light curves of O$_{03}$ and O$_{12}$ with a bin of 2000~s are extracted by setting the option
to be "Light curve".

The ephemeris from Equation \ref{equ:myminI} is used to calculate the phase of the X-ray light curve. In order to
correct the differences in the light arrival times due to the relative locations of the \textit{XMM-Newton}
and the star, the Modified Julian Date (MJD) of the X-ray light curve is converted
to HJD\footnote{http://www.physics.sfasu.edu/astro/javascript/hjd.html}. The bin of phase is eventually
set to 0.1 to get enough detection significance. We do not obtain the X-ray light curve from O$_{03}$ since
its effective time span is shorter than the orbital period of BH Cas. The phase-folded X-ray light curve of MOS of
O$_{12}$ is shown in Figure \ref{fig:LC}(c).

\begin{table*}
  \caption{ The \emph{XMM-Newton} observation log for BH Cas. Columns are organized as the observation ID and the define name of observation, observation instrument, the start and stop time of the observation, total exposure time and the effective exposure time.}
   \begin{center}
   \begin{tabular}{ccccccc}\hline \hline

Obs.ID & Def.Name  & Instrument & Start & Stop & Tot. Expo. & Effective Expo.  \\
       &          &            & (UT) & (UT) &  (ks)  &    (ks)   \\
0152260101 & O$_{03}$ & PN    & 2003-07-03$~$18:22:46 & 2003-07-04$~$06:30:01 & 43.635  & 33.6675   \\
0152260101 & O$_{03}$ & MOS1  & 2003-07-03$~$18:00:25 & 2003-07-04$~$06:34:57 & 45.272  & 36.7406   \\
0152260101 & O$_{03}$ & MOS2  & 2003-07-03$~$18:00:27 & 2003-07-04$~$06:35:01 & 45.274  & 36.7926   \\
0693390101 & O$_{12}$ & MOS1  & 2012-08-18$~$22:06:29 & 2012-08-20$~$10:14:43 & 130.094 & 100.495   \\
0693390101 & O$_{12}$ & MOS2  & 2012-08-18$~$22:06:29 & 2012-08-20$~$10:14:59 & 130.110 &	103.974   \\
\hline\noalign{\smallskip}
  \end{tabular}
  \end{center}
  \label{table:XMM-Newton_observation_log}
\end{table*}

\section{Data fitting} \label{sec:analysis}
\subsection{The O-C Analysis}\label{sec:O-C_analysis}

\begin{figure}
\centering
\includegraphics[width = 14cm]{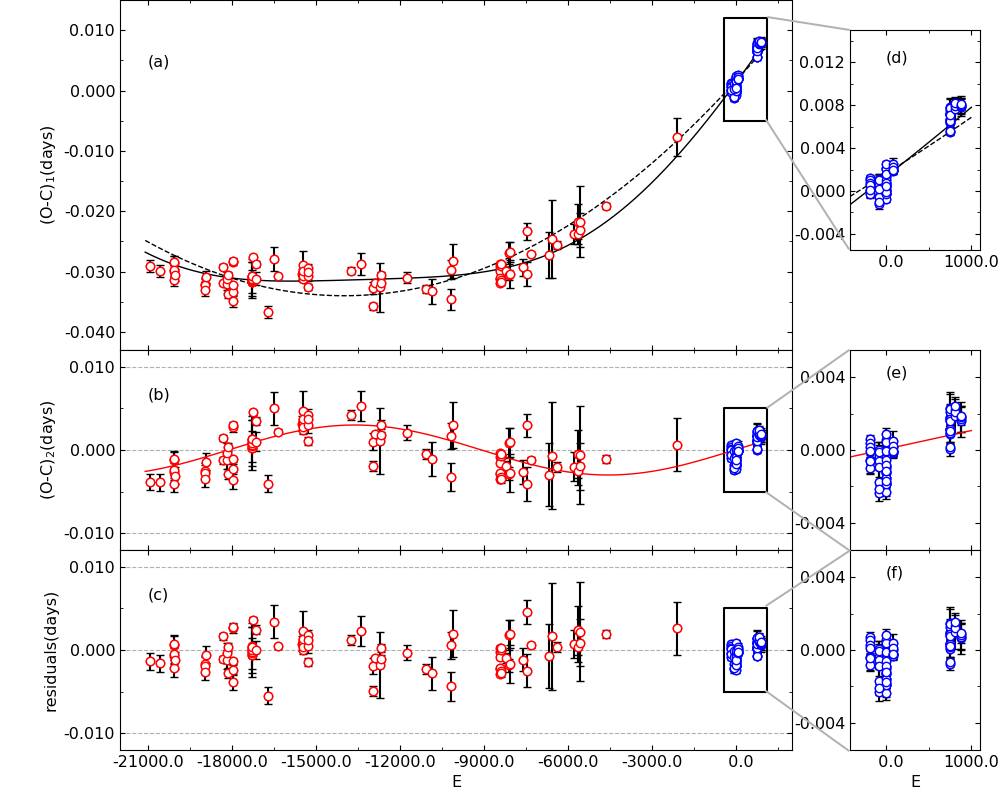}
\caption{(a): The $(\rm O-\rm C)_{1}$ diagram for BH Cas. Observed data are represented by red (literature) or blue (this work) points. The black dashed line is the parabolic least-squares fit (first three terms of Equation (\ref{equ:O_C_equation})) to the data. (b): The $(\rm O-\rm C)_{2}$ diagram (subtracting the dashed line from the observations in (a)). The red solid line is the fit to the data by $[a_{1}\,\textmd{cos} (\omega_{1}\,E)+b_{1}\,\textmd{sin}(\omega_{1}\,E)]$ with amplitude A$'$\,=\,0.00300\,days and oscillation period P\,$^{\prime}$\,=\,20.09\,yr and also plotted in panel\,(a) (black solid line) by adding to the black dashed line.(c): The residuals of panel(b). The distributions of our data at the ephemeral scale, in panel (a), (b) and (c), are zoomed in panel (d), (e) and (f), respectively.}
\label{fig:O_C}
\end{figure}

Ten years have passed since \citet{2009Obs...129...88A} pointed out that no evidence
had been found for any period change of BH Cas. We obtain 67 new minimum times in this work, combining with
94 minimum times collected from published papers and the website O$-$C Gateway
\footnote{http://var.astro.cz/ocgate/}, a total of 161 minimum times that cover 24 years from 1994 to 2018
are used for our O$-$C analysis. Equation~\ref{equ:myminI} is used to calculate the (O$-$C)$_{1}$
values. An obvious concave upward trend is evidenced in Figure \ref{fig:O_C}(a).
Therefore, we add a period derivative quadratic term to fit the data and the period increasing rate
is derived to be $dP/dt=+3.27 \times 10^{-7}$~d~yr$^{-1}$. The fit is statistically improved with a
reduced $\chi^2=34.3$. The corresponding residual $(O-C)_{2}$ is shown in Figure \ref{fig:O_C}(b).
The fitting remains poor, as $(O-C)_2$ seems to show a distinct quasi-sinusoidal oscillation
between epoch $-21000$ and $-11000$, whereas in the interval $-11000$ to 1000, the distribution is more or less
flat, with about 65\% of the data points below zero.
First, we consider the possibility of an elliptical orbit. There are no suitable solutions; either
the errors of the fitting parameters are too large, or the derived orbital eccentricity $e$ is greater
than 1, i.e., an non-ellipse orbit. Alternatively, we assume a circular orbit and use the following
formula \citep{2006AJ....132.2260H}:

\begin{equation}\label{equ:O_C_equation}
\begin{split}
(O-C)_{1}=&c_{0}\,+\,c_{1}\,\times\,E\,+\,c_{2}\,\times E^{2}\\
&\,+\, [a_{1}\,\textmd{cos} (\omega_{1}\,E)+b_{1}\,\textmd{sin}(\omega_{1}\,E)]\\
\end{split}
\end{equation}
where \textit{c$_{0}$}, \textit{c$_{1}$} and \textit{c$_{2}$} are the parabolic fitting parameters,
and \textit{a$_{1}$}, \textit{b$_{1}$} are the Fourier coefficients. The best-fit parameters are
listed in Table \ref{table:Period_changes} with the errors being directly output from the covariant matrix.
The revised quadratic ephemeris is then:

\begin{equation}\label{equ:Min_I}
\begin{split}
\textmd{{Min.I}}=&(\textmd{HJD$_{0}$})\,2458124.19660(25)\\
&\,+\,0.^{d}40589680(8)\,\times{E}\\
&+1^{d}.82(4)\,\times\,10^{-10}\,\times\,E^{2}.
\end{split}
\end{equation}

In Figure \ref{fig:O_C}(b), we use a red solid line to indicate the oscillation with the
amplitude $A'=0.00300$~days and the oscillation period $P'=2 \pi P/\omega=20.09$~yr.
 The residuals in Figure\ref{fig:O_C}\,(c) represent the deviation of data points in panel (b)
from the red solid line. The best-fit parameters of this oscillation are listed in Table \ref{table:Period_changes}. The reduced $\chi^{2}$ is 17.3\ in Figure \ref{fig:O_C}\,(c).

\begin{table}
  \caption{Best fit orbital parameters for BH Cas. }
  \label{table:Period_changes}
  \begin{center}
  \begin{tabular}{ccl}\hline \hline
$~~~~~$Date$~~~~~$&$~~~~~$Values$~~~~~$&$~~~~~$Error$~~~~~$ \\
\emph{c$_{0}$}               &  ~1.60\,$\times$\,10$^{-3}$  & ~2.5\,$\times$\,10$^{-4}$   \\
\emph{c$_{1}$}               &  ~5.09\,$\times$\,10$^{-6}$  & ~~~8\,$\times$\,10$^{-8}$   \\
\emph{c$_{2}$}               &  \,\,~1.82\,$\times$\,10$^{-10}$ & ~~~4\,$\times$\,10$^{-12}$   \\
\emph{a$_{1}$}               &  ~5.10\,$\times$\,10$^{-5}$                     & ~1.9\,$\times$\,10$^{-6}$   \\
\emph{b$_{1}$}               &  ~3.00\,$\times$\,10$^{-3}$                     & ~2.8\,$\times$\,10$^{-4}$   \\
\emph{$\omega_{1}$}              &  ~3.47\,$\times$\,10$^{-4}$                  & ~~~6\,$\times$\,10$^{-6}$   \\
\hline\noalign{\smallskip}
  \end{tabular}
  \end{center}
\end{table}

\subsection{Optical Light Curve Model Fitting}\label{sec:light curve model fitting}



The 2013 version of the Wilson-Devinney (W-D) code
(\citealt{1971ApJ...166..605W};\citealt{1979ApJ...234.1054W}; \citealt{2012AJ....144...73W}) is
used to fit the light curves in four bands in order to get the physical and geometrical parameters of BH Cas.
In the fitting process, we use the photometric data of this work and the radial velocity of
\citet{1999AJ....117.2503M} as the constraint condition. We define the primary component with subscript 1
and the secondary with subscript 2. In our fitting, the mass ratio $\textit{q}$ ($M_{2}/M_{1}$)= 0.475(23) obtained by radial velocity measurements \citep{1999AJ....117.2503M} is fixed through the fitting. The initial values of mean temperature of the primary (\textit{T}$_{1}$) and secondary (\textit{T}$_{2}$) are set to be 4850~K and 5300~K as mentioned in Section \ref{sec:Spectra}.
A circular orbit and synchronous rotation are assumed. For the bolometric and monochromatic limb
darkening law, we use the logarithmic form with the coefficients,
    \textit{X}$_{bolo}$, \textit{Y}$_{bolo}$, \textit{x}$_{B}$, \textit{y}$_{B}$, \textit{x}$_{V}$,
    \textit{y}$_{V}$, \textit{x}$_{R}$, \textit{y}$_{R}$, \textit{x}$_{I}$ and \textit{y}$_{I}$
    taken from \citet{1993AJ....106.2096V}.
    The gravity darkening coefficients and the bolometric albedos are set to
     \textit{g}$_{1}$\,=\,\textit{g}$_{2}$\,=\,0.32 \citep{1967ZA.....65...89L} and
     \textit{A}$_{1}$\,=\,\textit{A}$_{2}$\,=\,0.50 \citep{1969AcA....19..245R} because the
     atmospheres should both be convective. The adjustable parameters are: the orbit inclination \textit{i}, the mean surface temperature of primary \textit{T}$_{1}$,
      the mean surface temperature of secondary \textit{T}$_{2}$,
      the modified dimensionless surface potential \textit{$\Omega$}$_{1}$, and the bandpass luminosity of primary \textit{L}$_{1}$.

\begin{figure}
\centering
\includegraphics[width = 14cm]{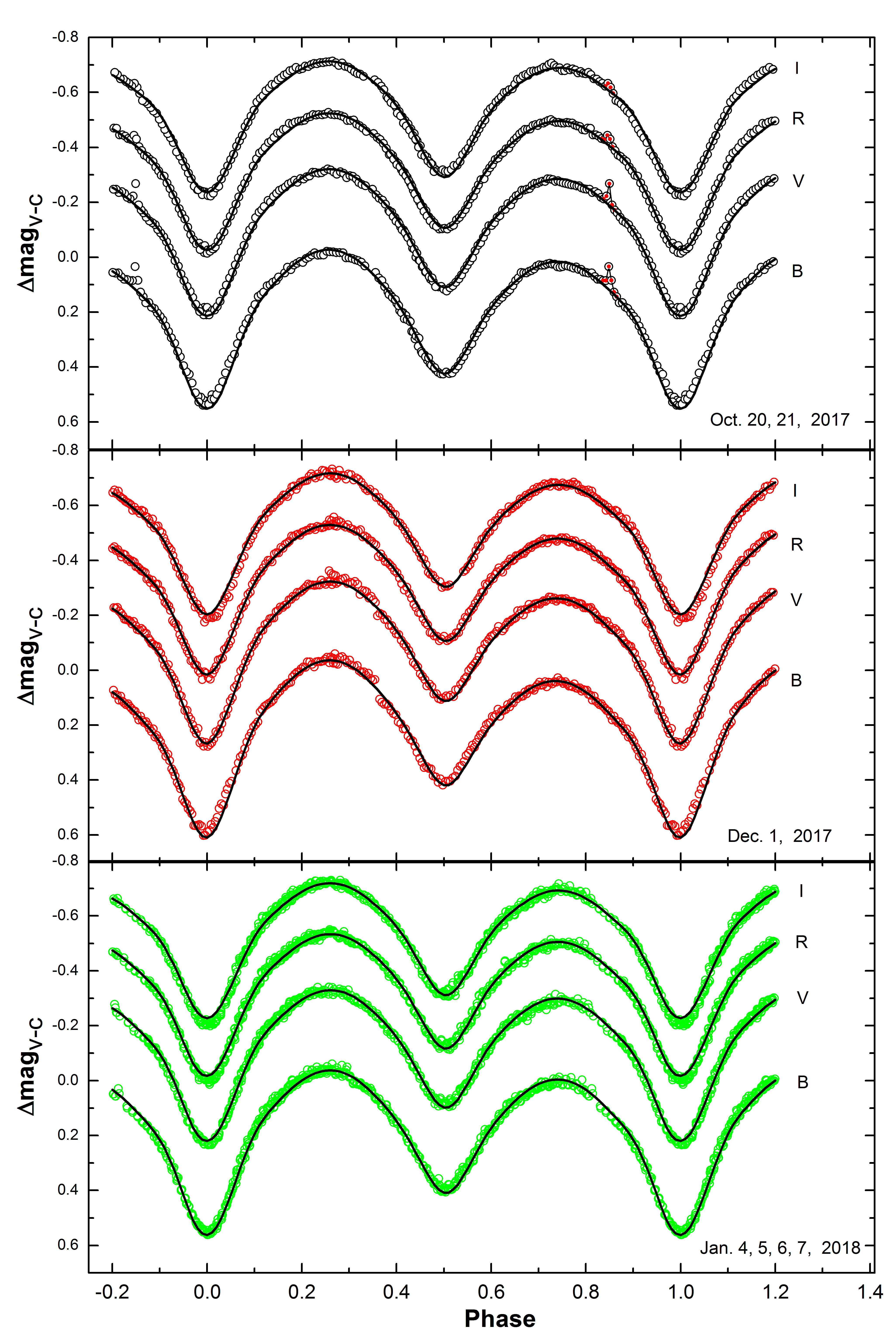}
\caption{Comparing of observed (open circles) and theoretical (solid lines) light curves in BVRI of BH Cas in different observations. The red solid points in panel (a) represent the optical flare. }
\label{fig:fitting_light_curve}
\end{figure}

\begin{figure}[ht]
\centering
\includegraphics[width = 16cm]{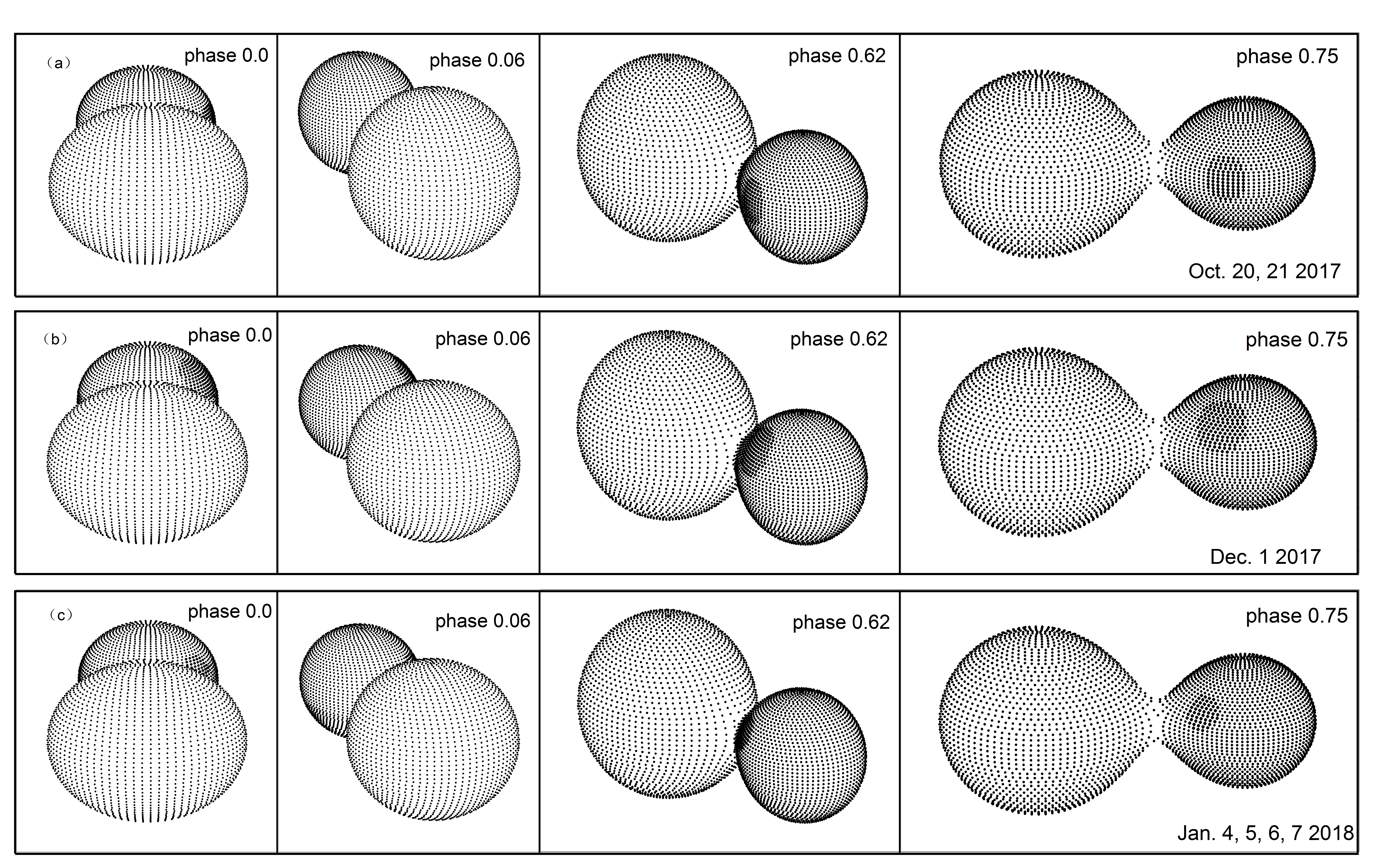}
\caption{The geometric configuration of BH Cas with a spot in the less-massive component at phases 0.00, 0.06, 0.62 and 0.75 in different observations.}
\label{fig:geometricconfiguration}
\end{figure}

As shown in Figure \ref{fig:O_C}(b), there is a cyclic modulation in the orbital period variation of BH Cas. So the third light \textit{L}$_{3}$ is taken into consideration in the light curves fitting. Because of the asymmetry of the light curves and the unequal luminosities at the maximum times
(see Figure \ref{fig:LC}), it is impossible to get perfect fitting to the light curves without star spots.
After a series of iterations we obtain fair solutions by adding a cool spot on the secondary star
by adjusting the longitude, latitude, angular radius and temperature of the spot. These solutions are
listed in Table \ref{table:WD_solution}. The parameters of BH Cas yielded by three sets of photometric
solutions are almost identical. In Figure \ref{fig:fitting_light_curve}, the theoretical light curves
given by the solution are compared to the observed ones in different observations.
Figure \ref{fig:geometricconfiguration} illustrates the geometric structure of BH Cas at different
orbital phases with a spot on the secondary component in different observations.
The revised absolute parameters we obtained for BH Cas are listed in Table \ref{table:Absolute_parameters}. We should note that the errors in Table \ref{table:WD_solution} and \ref{table:Absolute_parameters} are most likely underestimated \citep{2005ApJ...628..426P}. The reason is the existence of strong correlation between the fitted parameters due to due to the parameter space degeneracy of the model used.

\begin{table*}
  \caption{ \textit{BVRI} band photometric solutions of BH Cas obtained in October 2017, December 2017 and January 2018}
   \begin{center}
   \begin{tabular}{lcccccccc}\hline\hline
$~~~~~~~$Parameters &\underline{$~~~~~~$Best-fit Value(2017.10)$~~~~~~~$}    &\underline{$~~~~~~$Best-fit Value(2017.12)$~~~~~~~$}      &\underline{$~~~~~~$Best-fit Value(2018.01)$~~~~~~~$}\\
$~~~~~~~$           &Primary$~~~~~~~~~~~~~~~$Secondary                &Primary$~~~~~~~~~~~~~~~$Secondary                  &Primary$~~~~~~~~~~~~~~~$Secondary           \\
\hline\noalign{\smallskip}
$~~~~~~~$ $g1=g2$  (deg)        &  0.32                                                &  *                                           &  *                           \\
$~~~~~~~$ $A1=A2$  (deg)        &  0.50                                                &  *                                           &  *                                      \\
$~~~~~~~$ $X_{bolo} $       & 0.649   $~~~~~~~~~~~~~~~~~~~$  0.647           & *   $~~~~~~~~~~~~~~~~~~~$  *             & *   $~~~~~~~~~~~~~~~~~~~$  *           \\
$~~~~~~~$ $Y_{bolo} $       & 0.193   $~~~~~~~~~~~~~~~~~~~$  0.221           & *   $~~~~~~~~~~~~~~~~~~~$  *             & *   $~~~~~~~~~~~~~~~~~~~$  *  \\
$~~~~~~~$ $x_{B}$           & 0.847   $~~~~~~~~~~~~~~~~~~~$  0.829           & *   $~~~~~~~~~~~~~~~~~~~$  *             & *   $~~~~~~~~~~~~~~~~~~~$  *                \\
$~~~~~~~$ $y_{B}$           & 0.098   $~~~~~~~~~~~~~~~~~~~$  0.185           & *   $~~~~~~~~~~~~~~~~~~~$  *             & *   $~~~~~~~~~~~~~~~~~~~$  *                          \\
$~~~~~~~$ $x_{V}$           & 0.778   $~~~~~~~~~~~~~~~~~~~$  0.745           & *   $~~~~~~~~~~~~~~~~~~~$  *             & *   $~~~~~~~~~~~~~~~~~~~$  *                          \\
$~~~~~~~$ $y_{V}$           & 0.200   $~~~~~~~~~~~~~~~~~~~$  0.256           & *   $~~~~~~~~~~~~~~~~~~~$  *             & *   $~~~~~~~~~~~~~~~~~~~$  *                         \\
$~~~~~~~$ $x_{R}$           & 0.686   $~~~~~~~~~~~~~~~~~~~$  0.653           & *   $~~~~~~~~~~~~~~~~~~~$  *             & *   $~~~~~~~~~~~~~~~~~~~$  *                          \\
$~~~~~~~$ $y_{R}$           & 0.230   $~~~~~~~~~~~~~~~~~~~$  0.267           & *   $~~~~~~~~~~~~~~~~~~~$  *             & *   $~~~~~~~~~~~~~~~~~~~$  *                         \\
$~~~~~~~$ $x_{I}$           & 0.591   $~~~~~~~~~~~~~~~~~~~$  0.560           & *   $~~~~~~~~~~~~~~~~~~~$  *             & *   $~~~~~~~~~~~~~~~~~~~$  *                         \\
$~~~~~~~$ $y_{I}$           & 0.229   $~~~~~~~~~~~~~~~~~~~$  0.256           & *   $~~~~~~~~~~~~~~~~~~~$  *             & *   $~~~~~~~~~~~~~~~~~~~$  *                          \\
$~~~~~~~$ $i$  (deg)        &  71.51(85)                                              &  70.86(53)                             &  70.91(67)                              \\
$~~~~~~~$ $q$=$M_{2}/M_{1}$ &  0.475                                                    &  *                                        &   *                                      \\
$~~~~~~~$ $T$(K)            & 4899(124)   $~~~~~~~~~~~$   \,5109(139)          & 4862(64)   $~~~~~~~~~~~~~$   \,5214(65)        & 4874(66)   $~~~~~~~~~~~~~$   \,5163(74)      \\
$~~~~~~~$ $\Omega$          & 2.784(3)$~~~~~~~~~~~~~~~~$  2.784(3)          & 2.783(2)$~~~~~~~~~~~~~~~~$  2.783(2)      & 2.783(2)$~~~~~~~~~~~~~~~~$  2.783(2)                           \\
$~~~~~~~$ $L_{1}/(L_{1}+L_{2})_{B}$  &    0.589(3)                                      &    0.548(3)                              &    0.565(3)                \\
$~~~~~~~$ $L_{1}/(L_{1}+L_{2})_{V}$  &   0.605(3)                                       &   0.572(3)                               &    0.587(3)             \\
$~~~~~~~$ $L_{1}/(L_{1}+L_{2})_{R}$  &    0.618(3)                                      &    0.592(3)                              &    0.602(3)                \\
$~~~~~~~$ $L_{1}/(L_{1}+L_{2})_{I}$  &    0.625(3)                                      &    0.604(3)                              &    0.613(3)                              \\
$~~~~~~~$ $L_{3}/(L_{1}+L_{2}+L_{3})_{B}$  &   $2.18(9)\times10^{-3}  $                                 &  $2.04(6)\times10^{-3} $                             &    $2.26(8)\times10^{-3} $              \\
$~~~~~~~$ $L_{3}/(L_{1}+L_{2}+L_{3})_{V}$  &   $4.11(17)\times10^{-4} $                                 &  $3.92(9)\times10^{-4} $                             &    $4.32(10)\times10^{-4}$            \\
$~~~~~~~$ $L_{3}/(L_{1}+L_{2}+L_{3})_{R}$  &   $1.39(19)\times10^{-5} $                                 &  $1.64(13)\times10^{-5} $                             &   $1.52(11)\times10^{-5}$               \\
$~~~~~~~$ $L_{3}/(L_{1}+L_{2}+L_{3})_{I}$  &   $6.14(21)\times10^{-6} $                                 &  $5.74(17)\times10^{-6} $                             &   $5.93(19)\times10^{-6}$                \\
$~~~~~~~$ $r$(pole) & 0.4255(4)$~~~~~~~~~~~~~$ 0.3030(4)                    & 0.4287(2)$~~~~~~~~~~~~~$ 0.3063(2)        & 0.4275(3)$~~~~~~~~~~~~~$ 0.3051(3)           \\
$~~~~~~~$ $r$(side) & 0.4539(6)$~~~~~~~~~~~~~$ 0.3172(6)                    & 0.4581(3)$~~~~~~~~~~~~~$ 0.3213(3)        & 0.4566(4)$~~~~~~~~~~~~~$ 0.3198(4)       \\
$~~~~~~~$ $r$(back) & 0.4844(8)$~~~~~~~~~~~~~~$0.3547(9)                   & 0.4901(5)$~~~~~~~~~~~~~~$0.3614(4)        & 0.4880(5)$~~~~~~~~~~~~~~$0.3589(6)        \\
$~~~~~~~$ Latitude$_{spot}$(deg)  &  136.19(40.86)                                       &  101.11(2.03)                            &  120.27(31.70)                \\
$~~~~~~~$ Longitude$_{spot}$(deg) &  125.64(8.16)                                        & 115.98(89)                             &  86.70(14.41)          \\
$~~~~~~~$ Radius$_{spot}$(deg)    & 37.80(11.96)                                           & 41.12(73)                                & 27.39(48)      \\
$~~~~~~~$ $T_{spot}/T_{2}$        &  0.83(5)                                            &  0.80(1)                                 &  0.64(9)           \\
$~~~~~~~$ $f$(\%)                 & 15.14(4)                                              & 15.48(4)                                 & 15.48(4)              \\
\noalign{\smallskip}\hline
  \end{tabular}
  \end{center}
\label{table:WD_solution}
\tablecomments{ The asterisk (*) refers to assumed, fixed values.}
\end{table*}

\begin{table}
   \caption{Absolute parameters of BH Cas.}
   \begin{center}
   \begin{tabular}{lccccc}\hline \hline
Name&\textit{M}\,(M$_{\odot})$&\textit{$R$}\,(R$_{\odot})$&log\textit{\,g}\,(cgs)&\textit{$M_{bol}$}&\textit{$L$}\,(L$_{\odot})$\\
Primary&0.96(9) &   1.19 (7) &  4.27(6) & 5.01(7) & 0.703(13)\\
Secondary& 0.46(6) &   0.85(4) &  4.24(11) & 5.47(8) & 0.456(10) \\
\hline\noalign{\smallskip}
  \end{tabular}
  \end{center}

  \label{table:Absolute_parameters}
\end{table}

\subsection{X-Ray Spectra Fitting}\label{sec:Xray_spectra model fitting}

\begin{figure*}
  \centering
  \includegraphics[width = 18cm]{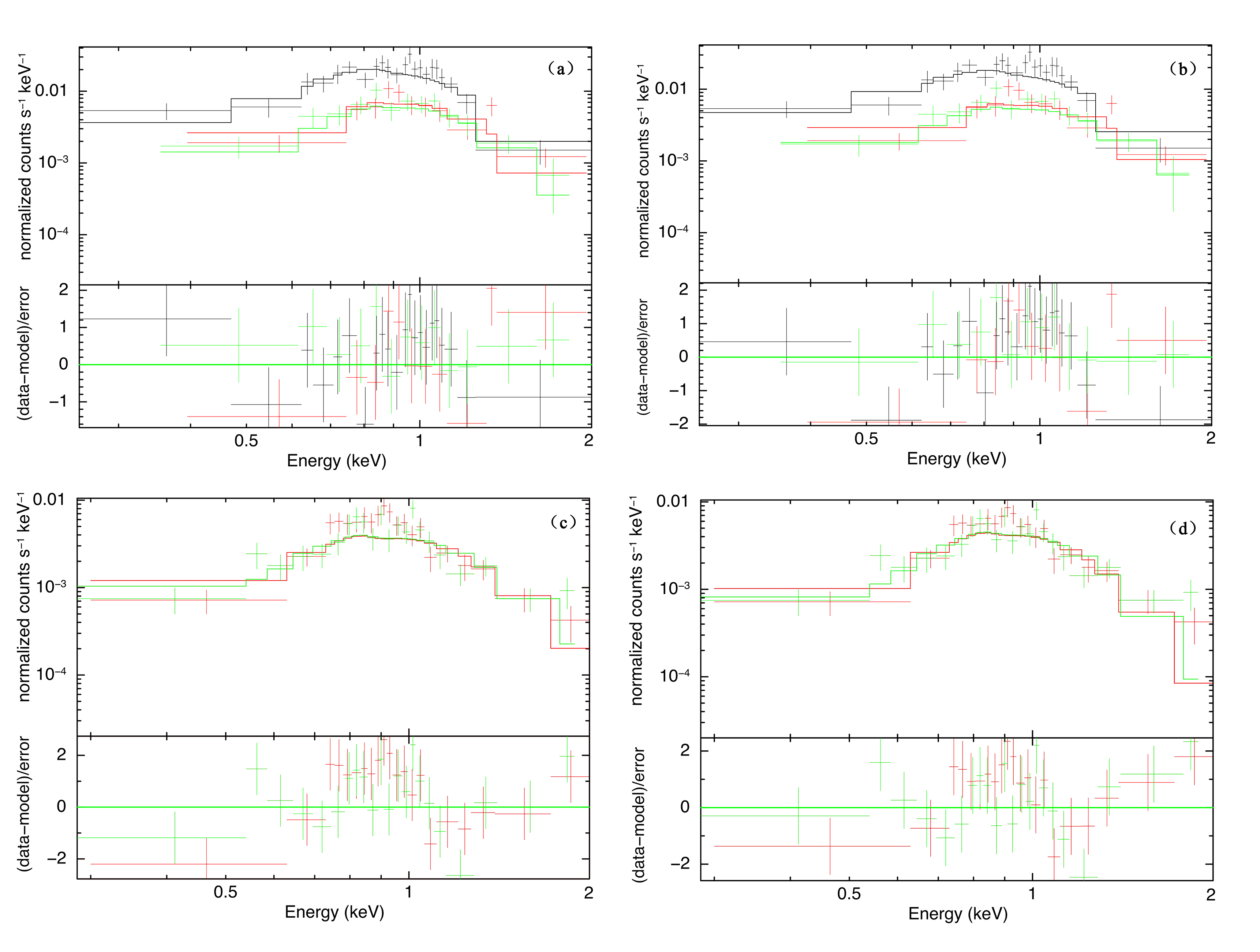}\\
  \caption{The X-ray spectra data (crosses) of BH Cas with different Obs.IDs and fitted by different radiation models (steps), PN colored in black, MOS1 colored in red and MOS2 colored in green. (a): Data are obtained in O$_{03}$ and fitted by black-body model. (b): O$_{03}$ and bremsstrahlung. (c): O$_{12}$ and black-body. (d): O$_{12}$ and bremsstrahlung. All of the lower panels represent the deviation of the data points from the fitting models.}\label{fig:015_PN_BB&Bremss_06_MOs2_BB&Bremss}
\end{figure*}

XSPEC version 12.10.0 is used to fit X-ray spectra. Since there are few photons above 2~keV, and X-ray emission is dominated by the
background, the fitting is carried out in the energy range 0.2--2~keV, i.e., the soft X-rays.
We separately try the thermal bremsstrahlung, black-body and the non-thermal power law to fit the
spectra, corresponding to \textit{bremss}, \textit{bbody} and \textit{powerlaw} radiative models in
XSPEC. We use the absorption model
\textit{wabs}\footnote{https://heasarc.gsfc.nasa.gov/xanadu/xspec/manual/node262.html}.
The average Galactic hydrogen column density \textit{$N_{H}$} in the region near BH Cas is estimated
to be $4.82 \times 10^{21}$~cm$^{-2}$ by the Leiden/Argentine/Bonn Survey \citep{2005A&A...440..775K}.
At first, we set \textit{$N_{H}$} as a free parameter. The lowest best fit valve is then
$6.07^{+1.94}_{-1.54}\,\times\,10^{21}$~cm$^{-2}$, which is not far from the derived value but
with too large an uncertainty.  We then fixed the value of \textit{$N_{H}$} as
$4.82\,\times\,10^{21}\,\textrm{cm}^{-2}$. We find that the power-law model failed to fit the
spectra, with all reduced $\chi^{2}\approx 2$. We separately carried out
simultaneous-fitting\footnote{https://heasarc.gsfc.nasa.gov/xanadu/xspec/manual/node39.html}
using \textit{bremss} or \textit{bbody} model for each observation. The reduced $\chi^{2}$
values in O$_{12}$ are all greater than those in O$_{03}$, which is most likely caused by the
exposure times and the instrumental differences between MOS and PN, e.g., quantum efficiency
and time resolution \citep{2001A&A...365L..18S, 2001A&A...365L..27T}.
Fitting parameters are listed in Table \ref{table:best_fit_X_ray}, and the spectra with the
corresponding best-fit models are shown in Figure \ref{fig:015_PN_BB&Bremss_06_MOs2_BB&Bremss}.
For the black-body and the bremsstrahlung models of O$_{03}$, the plasma temperatures
$\textit{kT}_{\textrm{03}}$ are $0.137^{+0.006}_{-0.006}$~keV and
$0.266^{+0.020}_{-0.017}$~keV, respectively, and for O$_{12}$, they are
$0.132^{+0.006}_{-0.006}$~keV and $0.258^{+0.019}_{-0.016}$~keV.
The derived X-ray flux in the energy range 0.2--2~keV is estimated to be
$\thicksim3.77^{+0.27}_{-0.28}\,\times\,10^{-14}$\,erg\,cm$^{-2}$\,s$^{-1}$ in 2003,
and $\thicksim4.16^{+0.21}_{-0.21}\,\times\,10^{-14}$\,erg\,cm$^{-2}$\,s$^{-1}$ in 2012.

\begin{table*}
  \caption{The best fitting parameters for black-body and bremsstrahlung models for the X-ray spectra of BH Cas.}
   \begin{center}
   \begin{tabular}{cccc}\hline\hline
$~~~~~~~$Obs.ID$~~~~~~~$  &Parameters & Black Body & Bremsstrahlung \\
       0152260101      & $kT\,(\textrm{keV})$  &$0.137^{+0.006}_{-0.006}$& $0.266^{+0.020}_{-0.017}$\\
                           &Normalization &$6.50^{+1.09}_{-0.97}\,\times\,10^{-6}$&$1.91^{+0.51}_{-0.41}\,\times\,10^{-3}$\\
                           & Reduced $\chi^{2}$ (d.o.f\,\footnote{the degree of freedom})   &1.06(74)                            & 1.25(74)\\
                           & Absorbed Flux&        $3.72^{+0.37}_{-0.38}$   &    $3.82^{+0.39}_{-0.40}$                         \\
                           & \,$(\times\,10^{-14}\textrm{erg}\,\textrm{cm}^{-2}\,\textrm{s}^{-1})$\\
       0693390101          & $kT\,(\textrm{KeV})$  &$0.132^{+0.006}_{-0.006}$& $0.258^{+0.019}_{-0.016}$\\
                           &Normalization &$7.95^{+1.33}_{-1.18}\,\times\,10^{-6}$&$2.27^{+0.62}_{-0.51}\,\times\,10^{-3}$\\
                            & Reduced $\chi^{2}$ (d.o.f)   &1.55(58)                           & 1.84(58)\\
                           & Absorbed Flux&        $4.12^{+0.29}_{-0.30}$   &         $4.20^{+0.31}_{-0.29}$                     \\
                           & \,$(\times\,10^{-14}\textrm{erg}\,\textrm{cm}^{-2}\,\textrm{s}^{-1})$\\
\noalign{\smallskip}\hline
  \end{tabular}
  \end{center}
\label{table:best_fit_X_ray}
\end{table*}

\section{Discussions}\label{sec:Discu}

\subsection{Additional Companion}

The cyclic oscillation shown in Figure \ref{fig:O_C}\,(b) may be caused by the Applegate
mechanism \citep{1992ApJ...385..621A}, or by the light-time effect as a consequence of
the existence of a third body. By calculating the quadrupole moment of the solar-type components,
we can quantify the role played by the Applegate mechanism. The equation given by \citet{2000A&A...354..904R}:
\begin{equation}\label{equ:Applegate P}
    \Delta P = \sqrt{2\,[1-\textmd{$\cos$}(2\pi P/P^{\prime})]} \times A,
\end{equation}
where \textit{A} is the semi-amplitude, \textit{P\,$^{\prime}$} is the oscillation period , and $\textit{A}\,=\,\sqrt{a_{1}^{2}+b_{1}^{2}}$\,=\,0.00300 day. By using the relationship between \textit{$\omega$} and \textit{P\,$^{\prime}$},
    \textit{$\omega$}\,=\,2\textit{$\pi$}\textit{P/P\,$^{\prime}$}, we get the value of \textit{P\,$^{\prime}$} in year. So the periodic rate of change is \textit{$\Delta$P/P}\,$\sim$\,2.568\,$\times$ 10$^{-6}$. To produce such a change rate, the change in the required quadrupole moment
can be calculated using Equation \citep{2002AN....323..424L}:

\begin{equation}\label{equ:Applegate dP/p}
    \Delta P/P\,=\,-9\,\frac{\Delta Q}{M a^{2}},
\end{equation}
where \textit{M} is the mass of each component in solar mass, and the semi-major axis \textit{a} is the distance between two components. \textit{a} can be solved using Kepler's third law:
\begin{equation}\label{equ:Kepler third law}
    M_{1} + M_{2} = 0.0134 \frac{a^{3}}{P^{2}}.
\end{equation}

Combining Equations \ref{equ:Applegate dP/p} and \ref{equ:Kepler third law},
we then obtain the required quadrupole momentums variations of
\textit{$\Delta$Q$_{1}$}\,$\sim$\,1.8\,$\times$\,10$^{49}$\,g cm$^{2}$ and
\textit{$\Delta$Q$_{2}$}\,$\sim$\,8.8\,$\times$\,10$^{48}$\,g cm$^{2}$.
However, these are less than the range of typical values for a binary system with periodic oscillations
that can be explained by this mechanism is 10$^{51}$~g~cm$^{2}$ to 10$^{52}$~g~cm$^{2}$
\citep{2002AN....323..424L}. In addition, this mechanism is mainly used to explain period modulations of an amplitude \textit{$\Delta$P/P}\,$\sim$10$^{-5}$\citep{1992ApJ...385..621A} while this value of BH Cas is $\sim$10$^{-6}$. Therefore, we conclude that the Applegate mechanism is unable to explain the periodic change of this source. Alternatively, we consider the light-time effect (LITE) to
explain the cyclic oscillation. By combining the orbit parameters listed in Table \ref{table:Period_changes},
we get the parameters of the third body using the following formula \citep{1978ASSL...68.....K}:

\begin{equation}\label{equ:a12sini}
    a\,'\,\textmd{sin}i'=\textmd{c}\sqrt{a_{1}^{2}+b_{1}^{2}},
\end{equation}

where $\textit{a}\,'$ is the orbital radius of the common center of mass of the triple system
consisting of the contact binary and a third body, and \textit{i\,$'$} is the inclination of
the third body. 
We further calculate the mass of the third object \textit{$M\,'$} and the orbital radius \textit{$a\,'$} via:

\begin{table*}
   \caption{Derived parameters of the third body of BH Cas. }
   \begin{center}
   \begin{tabular}{lc}\hline \hline
$~~~~~~~$Parameters &\underline{$~~~~~~~$Third Body$~~~~~~~$}  \\
$~~~~~~~$ & $~~~~~$Values$~~~~~~~~~$Error$~~~~~$                  \\
\hline
\emph{P\,$'$}\,(yr)                &  $~~~~~$20.09$~~~~~~~~~~\,~$0.14   \\
\emph{a\,$'$}sin\cite{}\emph{i\,$'$}\,(AU)   &  $~~~~~$0.52$~~~~~~~~~~~~~$0.06 \\
\emph{f\,$\rm (\textit{M}'\rm)$}\,(M$_{\odot}$) & $~~$3.48\,$\times$\,10$^{-4}$$~~~$2.6\,$\times$\,10$^{-5}$   \\
\emph{M$\,'\,$\rm(i\,$'$\,=\,90\,$^{\circ}$\rm )}\,(M$_{\odot}$)   & $~~~~$ 0.093$~~~~~~~~~~~$0.007  \\
\emph{a$\,'\,$\rm(i\,$'$\,=\,90\,$^{\circ}$\rm )}\,(AU)   &  $~~~~~$7.94$~~~~~~~~~~\,~~$0.13 \\
\hline\noalign{\smallskip}
  \end{tabular}
  \end{center}
  \label{table:third_fourth_parameters}
\end{table*}

\begin{equation}\label{fm3}
    f\,\rm (\textit{M}\,'\rm)\,=\,\frac{(\textit{M}\,'\sin\textit{i}\,')^{3}}{(\textit{M}_{1}+\textit{M}_{2}+\textit{M}\,')^{2}}\,=\,\frac{4\pi^{2}}{GP^{\prime2}}(\textit{a}\,'\,\sin\textit{i}\,')^{3},
\end{equation}
\begin{equation}\label{a3}
    d\,'\,=\,\frac{M_{1}+M_{2}}{M\,'}a\,',
\end{equation}
where G is the gravitational constant. The computed astrometric orbit makes the existence of the LITE evident. The mass of the third object thus derived
is 0.098(9)\,M$_{\odot}$ if its orbit is coplanar (\textit{i}\,=\,71$^{\circ}$) with the binary system. The derived properties of the third body are listed in Table \ref{table:third_fourth_parameters}. In addition, the light curve fitting to our observations (see Table \ref{table:WD_solution}) shows that the contribution of the third light to the total luminosity reaches two thousandths in B band, which may also support the possible existence of a third body.

\subsection{The Magnetic Field Activity}\label{sec:Mageticfieldactivity}

As shown in Table~\ref{table:maximum_magnitude}, all the magnitudes of maximum times around phase 0.25 are
significantly higher than those around 0.75. We thereby conclude that, for the first time, we detect significant O'Connell effect in \textit{BVRI} bands for BH\,Cas. As shown in Figure \ref{fig:LC}(a), we note that the light curves in all bands scatter noticeably
from phase 0.6 to 1.06 comparing to that from phase 0.06 to 0.6, and the magnitude difference between the amplitude of the primary maxima and the following secondary changes from season to season.
An optical flare was detected on Oct. 21, 2017, and this is the first detection of an optical flare
in this source. The variations in the $B$ and $V$ bands, which amount to $\sim0.15$~mag and
continue for $\sim$15 minutes, show the flare near phase 0.85. In R and I bands the tendency also can be
recognized. The luminosity of the flare decreases as the wavelength increases, similar to the situation in
U Pegasi \citep{1952PASP...64..200H}.
 One possible explanation for the O'Connell effect is the magnetic activities,
manifest by flares \citep{2012MNRAS.423.3646Q,2014ApJS..212....4Q} or star spots \citep{1997ASPC..130..129M}.
But flares are sporadic and short in duration, hence, difficult to account for the persistence of the O'Connell
effect. However, star spots are relatively long-lived, and their
existence and the distribution on the surface can be diagnosed with photometric observations at different
rotational phases. By adding a cool spot on the less massive component, the W-D code fitting leads to
a good result for the different luminosity of the maximum times and the asymmetry in optical light curves.
In three observation seasons, the variation of the amplitude of the light curves is likely caused by the
changes of the total spot area, location, and temperature, i.e. strong spot activities. From the W-D solution, the spot began to
appear at phase $\sim$0.6, was completely exposed at phase $\sim$0.75, and was covered by the primary
star at phase $\sim$1.06, which are in good agreement with the observations. These variations in the
spot of the secondary suggest a stronger magnetic activity than that of the primary.
No flare and significant O'Connell effect have been reported in previous works for BH Cas \citep{1999AJ....117.2503M,2001A&A...374..164Z,2001hell.confE..77N}.
 Our detections suggest that this source in its stronger magnetic active state during our observations than that of previous observations. In addition, there are shallow absorptions in the H$\alpha$, H$\beta$, H$\gamma$ \citep{2017AJ....154..260P} and Ca II triplet lines \citep{1997A&AS..124..359M} in the spectra of Figure \ref{fig:spectra_matchK3V} and \ref{fig:spectra_matchG8V}. These may indicate that BH Cas is chromospherically active.

\subsection{The Spectral Types}

Our low-resolution spectra show that the spectral types at phase $\sim$0.0 and $\sim$0.5 are K3V$\pm$1 ($\sim$4850$\pm$150\,K) and G8V$\pm$2 ($\sim$5300$\pm$\,300\,K), respectively. As shown in Table \ref{table:WD_solution}, the temperatures of the primary $T_{1}$ and secondary $T_{2}$ obtained from light-curve fitting are $\sim$4880$\pm$100K ($\sim$K3V$\pm$1) and $\sim$5160$\pm$100\,K ($\sim$G9V$\pm$1), respectively. Using the inclination \textit{i} in Table \ref{table:WD_solution} and absolute parameters listed in Table \ref{table:Absolute_parameters}, the geometric model of BH Cas is simplified as two spheres with external tangent. If we do not consider the limb darkening effect, the ratios of luminosities of primary to secondary are derived to be $\sim$7:1 at phase 0.0 and $\sim$1:1 at phase 0.5. This may indicate that the temperature of the primary should be slightly lower than the spectral temperature at phase 0.0 while the temperature of secondary higher than the spectral temperature at phase 0.5. Considering the errors of the light curve fitting are most likely underestimated, the fitting temperature of primary is close to the spectral temperature at phase 0.0, which may be caused by the dominant luminosity of the primary at this phase and may suggest that the spectral type of primary is likely to be $\sim$K3V$\pm$1. According to the light curve fitting, the spectral type of the secondary is $\sim$G9V$\pm$1, while the photometric contribution and the spectral type at 0.5 phase indicated that the temperature of secondary is higher than the spectral temperature of G8V$\pm$2. The inconsistency between these results makes it difficult for us to judge the actual spectral type of the secondary. Considering the magnetic activity of BH Cas mentioned in Section \ref{sec:Mageticfieldactivity}, the spectrum may be influenced by stellar activities, e.g. spots, flares, resulting in deviation of spectrum temperature from typical stellar temperature of BH Cas.

\citet{1999AJ....117.2503M} derived an effective temperature for BH Cas about 4600$\pm$400\,K by the $V-I$ and $R-I$ colors. This temperature corresponds to a main-sequence spectral type K4$\pm$2. Using W-D code, he derived the temperatures of the primary and secondary star as 4790$\pm$100K and 4980$\pm$100K, which are close to the fitting temperatures of this work. A temperature of BH Cas released by Gaia2 is 4990$\pm$324\,K\citep{2018A&A...616A...8A}. Although the uncertainty of the temperature is high, it is within the range of component temperature of BH Cas we derived by spectra. Based on a single spectrum taken in the range 5080\,-\,5290~\AA with a resolution of 0.2\,\AA/pixel using a 1.9-m telescope \citep{1999AJ....118..515L}, \citet{2001A&A...374..164Z} set the temperature of secondary component to 6000K according to spectral type of F8($\pm$2). By fixing $T_{2}$ to 6000\,K, $T_{1}$=5500\,K is obtained using W-D code. The spectral type of F8($\pm$2) is significantly different from what we got at phase of 0.0 or 0.5. Although a strong stellar activity could be a possible reason, we may suggest that such a difference is most possibly due to a high uncertainty in the determining the spectral type using such a narrow range (210\AA) and high resolution. More low resolution spectral observations at different phases may help us to explore component spectral types of BH Cas.

\subsection{The X-ray Emission}

While the variation in the optical light curves is caused by the binary geometry and inclination,
the X-ray light curve does not show obvious eclipses. The lack of X-ray variability, with a
reduced $\chi^2$ of the phase-folded X-ray light curve (Figure \ref{fig:LC}(c)) being, 1.5, could be
the consequence of insufficient number of X-ray photons. Alternatively this indicates a different
light distribution of X-ray from that of the optical one. The fitting of asymmetry and O'Connell effect
in the optical light curves suggests the presence of spot. The X-ray emission may be caused by the
chromosphere and corona activity \citep{ 2006ApJ...650.1119H, 2016AJ....151..170H, 2015MNRAS.446..510K}.
The location, size and temperature of the spot change with time. This change may be reflected in the
X-ray light curve or spectra. However, X-ray and optical observations are not carried out at the same time.
Simultaneous observations of X-ray and optical are necessary to solve the issue.
Both of the single models, the black-body and the bremsstrahlung, can describe the X-ray spectra of
BH Cas, which indicates that the origin of X-ray emission of this source is thermal.
The derived fluxes in the energy range 0.2--2~keV are estimated to be
$\thicksim3.77^{+0.27}_{-0.28}\,\times\,10^{-14}$\,erg\,cm$^{-2}$\,s$^{-1}$ in 2003,
and $\thicksim4.16^{+0.21}_{-0.21}\,\times\,10^{-14}$\,erg\,cm$^{-2}$\,s$^{-1}$ in 2012.
Using the Gaia2 parallax $2.2646\pm0.0288$~mas \citep{2018yCat.1345....0G}, the distance
of BH Cas is estimated to be 441$^{+6}_{-5}$ pc. The average X-ray luminosities of BH Cas are then
estimated as $L_{03}\,=\,8.76^{+0.63}_{-0.65}\,\times\,10^{29}$\,erg\,s$^{-1}$ and
$L_{12}\,=\,9.67^{+0.34}_{-0.49}\,\times\,10^{29}$\,erg\,s$^{-1}$. In 1997, the X-ray flux of BH Cas
in 0.1--2.5~keV is measured to be $4.2\,\times\,10^{-14}$\,erg\,cm$^{-2}$\,s$^{-1}$ \citep{1997MNRAS.291..709B}
 and the average X-ray luminosity in this observation is about $9.67\,\times\,10^{29}$\,erg\,s$^{-1}$.
Although the X-ray flux in the light curve fluctuates between different phases of the complete phase, there is no significant change
 in these average X-ray luminosities, which indicates that the average X-ray luminosity of BH Cas may be stable over long time.

\subsection{Mass Transfer and Evolution of Angular Momentum}

The solution of the W-D code, revealing a hotter less-massive component, indicates that BH Cas is a
W-subtype W UMa system. Moreover, the O$-$C analysis shows a long-term increase in the orbital period
which can be explained by the transfer of matter from the less-massive component to the more-massive one.
Using the following equation \citep{1991MNRAS.253....9T},

\begin{equation}\label{equ:masstransfer}
    \frac{\dot{P}}{P}=3\dot{M_{1}}(\frac{1}{M_{2}}-\frac{1}{M_{1}}),
\end{equation}
where the mass of the two components, \textit{M$_{1}$}\,=\,0.96\,M$_{\odot}$,
\textit{M$_{2}$}\,=\,0.46\,M$_{\odot}$, we derive the mass transfer rate
\textit{d$M_{1}$/dt}\,=\,+2.37\,$\times$\,10$^{-7}$ M$_{\odot}$/yr. The plus symbol means that
the more massive component is gaining mass from the less massive star
at a rate of \textit{dM$_{2}$/dt}\,=\,$-$2.37\,$\times$\,10$^{-7}$\,M$_{\odot}$/yr.
The mass transfer time scale is then
$\textit{$\tau$}$\,$\sim$\,\textit{M$_{2}$/$\dot{M_{2}}$} $\sim$\,1.48 \,$\times$\,10$^{6}$ yr.
The coupling of this mass transfer process with the increase of orbital period is consistent with
the prediction of the angular momentum loss (AML) theory (e.g., \citet{2000Ap&SS.271..331S}).

\begin{figure*}
\centering
\includegraphics[width = 14cm]{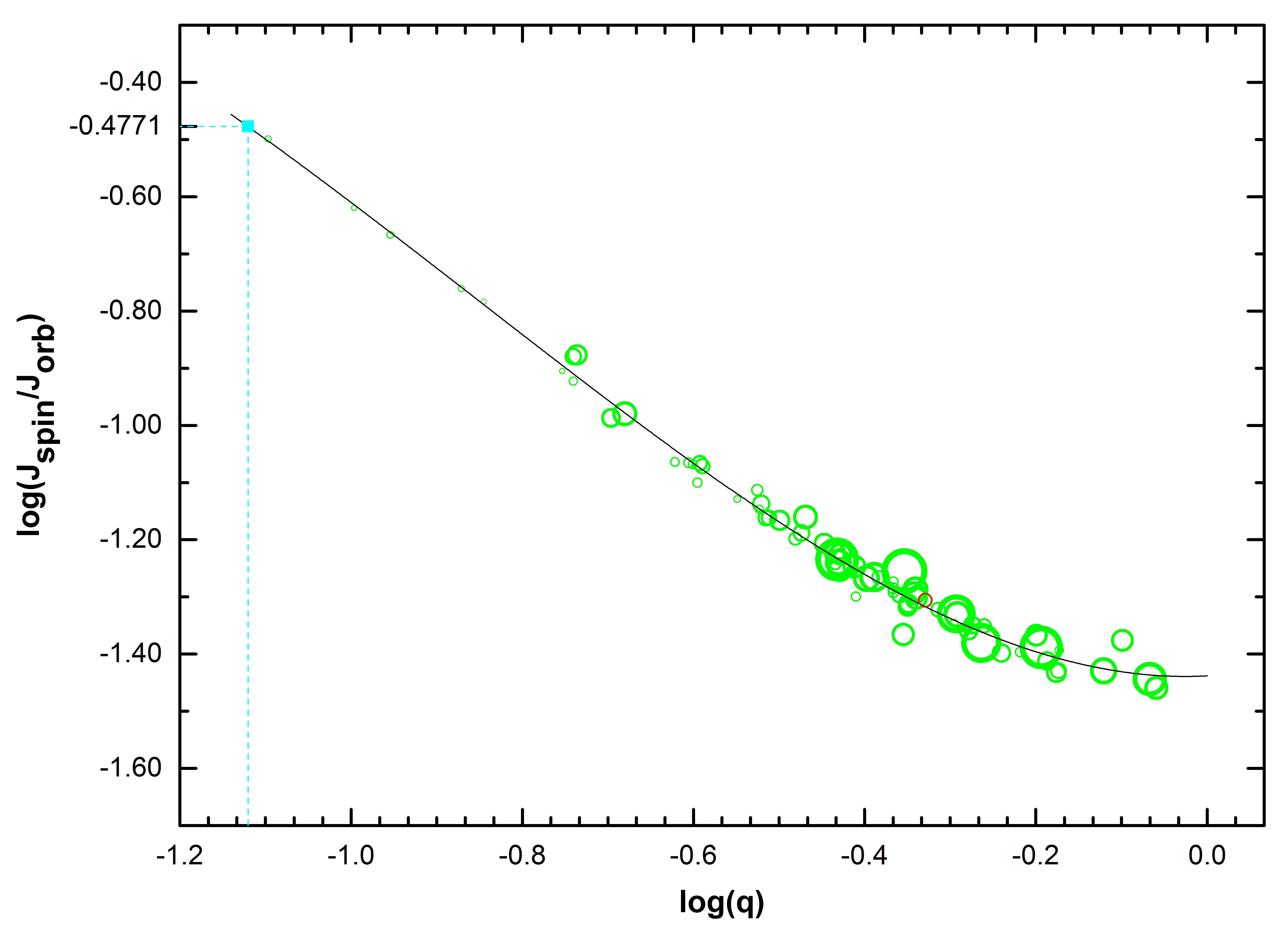}
\caption{Plots of $log(q)$\,-\,$log(J_{spin}/J_{orb})$ of W-subtype contact binaries. The least-square fitting line is a solid black line. The size of the circle corresponds to the total angular momentum. The cyan square point is the intersection of the black curve with the horizontal line y\,=\,log(1/3). BH Cas is signed by the red open circle.}
\label{fig:AMN}
\end{figure*}

The orbital angular momentum and the spin angular momentum \citep{2007MNRAS.377.1635A} of
binaries can be estimated by

\begin{equation}\label{equ:Jorb}
    J_{orb}\,=\,\frac{M_{1}M_{2}}{M_{1}+M_{2}}a^{2}\Psi,
\end{equation}

\begin{equation}\label{equ:masstransfer2}
    J_{spin}\,=\,k_{1}^{2}M_{1}R_{1}^{2}\Psi_{1}+k_{2}^{2}M_{2}R_{2}^{2}\Psi_{2},
\end{equation}
where \textit{$\Psi_{i}$s} are the orbital angular velocities, $k_{i}^{2}s$ are the dimensionless
gyration radii, and $R_{i}s$ are the volume radii. We can assume synchronization for close binaries,
so \textit{$\Psi_{1}$}\,=\,\textit{$\Psi_{2}$}\,=\,2$\pi$\,/\,\textit{P}. Since the possible evolutionary
relationship between W-subtype and A-subtype is still controversial, it is reasonable to study their
angular momentum evolution separately. In order to investigate the angular momentum of W-subtype binaries,
according to their period and total mass, i.e., \textit{M}$_{1}$\,+\,\textit{M}$_{2}$\,$<$\,3,
and \textit{P}\,$<$\,1day, we compile a list of 73 objects for which the absolute parameters have been well determined.
Their basic parameters are listed in Table \ref{table:A-W-subtype}, sorted by the decreasing total mass.
We derive the values of J$_{spin}$/J$_{orb}$ and J$_{spin}$\,+\,J$_{orb}$ for all these binaries.
After a series of least-squares fitting in different forms, we obtain the best fitting curves
as shown in Figure \ref{fig:AMN}. The equation between $\log(J_{spin}/J_{orb})$ and $\log(q)$ is as follows:

\begin{equation}\label{equ:masstransfer3}
\begin{split}
\log(J_{spin}/J_{orb})\,=&\,-1.43817-0.08274\times \log(q)\\
&+1.63025\times(\log(q))^{2}\\
&+0.83308\times(\log(q))^{3}\\
&+0.11355\times(\log(q))^{4}.
\end{split}
\end{equation}

In Figure \ref{fig:AMN}, the green open circles represent the W-subtype W UMa-type contact binaries,
and the fitting curve is in black, with the size of each circle corresponding to the value of the angular
momentum. In general, as $\log(q)$ decreases, $\log(J_{spin}/J_{orb})$ increases.
The cyan square point is the position where $J_{spin}/J_{orb}$\,=\,1/3,
we obtained minimal mass ratio $q_{min,W-subtype}\,=\,0.0757(4)$. This value of the secular tidal instability
is consistent with the range 0.076---0.078 reported by \citet{2006MNRAS.369.2001L}, and is marginally
higher than the range 0.070--0.074 presented by \citet{2009MNRAS.394..501A}. Adopting the absolute
parameters we obtained for BH Cas, $J_{spin}/J_{orb} =0.04531$.
Due to the total mass of BH Cas is small, the angular momentum of BH Cas is smaller than
other source around it (see Figure \ref{fig:AMN}). With the transfer of mass between the two components of BH\,Cas,
the mass ratio will decrease, while $J_{spin}/J_{orb}$ increases. If the system reaches a state
when $J_{\textrm{orb}}$\,=\,3\,$J_{\textrm{spin}}$ \citep{2007MNRAS.377.1635A}, tidal instability will
occur, thereby forcing the binary to merge into a single, rapidly rotating star.

\section{Summary}\label{sec:Summary}

We present the new photometric and optical spectral observations of BH Cas.
The O$-$C analysis and the theoretical solution of optical light curves indicate BH Cas to be
a W-subtype W UMa contact binary with an orbital period increased with a rate of
$\textit{dP/dt}\,=\,+3.27 \times 10^{-7} \textrm{d}\,\textrm{yr}^{-1}$.
The change of the orbital period may be caused by the mass transfer from $M_{2}$ to $M_{1}$. One cyclical oscillation ($A'=0.00300$~days and $P'=20.09$~yrs) superimposed on the long-term period increasing tendency,
 which we attribute to a additional components in the system,
e.g., a low-mass star.
In the multi-band optical light curves, the strong spot activities and
the optical flare are detected at the same time on a W UMa-type system for the first time.
Solutions using the W-D code suggest the existence of a cool spot on the secondary star.
The spectra observed at phases $\sim$0.0 and $\sim0.5$ in 2018 and 2019 led to spectral typing of
K3V$\pm1$ and G8V$\pm2$, respectively. The primary star likely has a spectral type of K3V$\pm1$.
The absolute parameters of BH Cas are obtained, including the mass, radius, and surface gravity, etc.
The X-ray data extracted from the \textit{XMM-Newton} show a light curve with no evidence of occultation
as seen in the optical, and X-ray spectra consistent with black-body or bremsstrahlung thermal emission.

\section{Acknowledgements}\label{sec:Acknow}

We thank the anonymous referee very much for his/her constructive suggestions, which helped to improve the paper a lot.
We acknowledge the support of the staff of the Xinglong 2.16m telescope. This work was partially supported by the Open Project of the Key Laboratory of Optical Astronomy, National Astronomical Observatories, Chinese Academy of Science.
We also acknowledge the support of X-ray data based on observations obtained with \textit{XMM-Newton}, an ESA science mission with instruments and contributions directly funded by ESA Member States and NASA.
This research is supported by the National Natural Science Foundation of China (grant Nos.11873081, 11661161016), the program of the light in Chinese Western Region (LCWR; grant Nos.2015-XBQN-A-02),2017 Heaven Lake Hundred-Talent Program of Xinjiang Uygur Autonomous Region of China, the Youth Innovation Promotion Association CAS (grant Nos.2018080). We thanks to Prof. Wenping Chen and Weiyang Wang, for their comments on this paper.

\appendix

\section{BASIC Parameters of W UMa type and Near-contact binaries}
\startlongtable
\begin{deluxetable*}{lccccccccl}
\tablecaption{Fundamental properties of W-subtype W UMa type binaries. Columns are organized as name, subtype, period, radii of primary and secondary, primary and secondary masses, the ratio of spin specific angular momentum to orbital specific angular momentum, the sum of spin specific angular momentum to orbital specific angular momentum and the numbers for references.\label{table:A-W-subtype}}
\tabletypesize{\scriptsize}
\tablehead{
\colhead{Name} & \colhead{Subtype$~~~~$} &
\colhead{$~~~~$Period$~~~~$} & \colhead{$~~~~$R$_1$$~~~~$} & \colhead{$~~~~$R$_2$$~~~~$} &
\colhead{$~~~~$M$_1$$~~~~$} & \colhead{$~~~~$M$_2$$~~~~$} &
\colhead{$~~~~$J$_{spin}$/J$_{orb}$$~~~~$} & \colhead{$~~~~$J$_{spin}$+J$_{orb}$$~~~~$} &
 \colhead{$~~~~$Ref$~~~~$} \\
\colhead{} & \colhead{} &
 \colhead{d} & \colhead{(R$_{\odot}$)} & \colhead{(R$_{\odot}$)} &
\colhead{(M$_{\odot}$)} & \colhead{(M$_{\odot}$)} &
\colhead{} & \colhead{}
& \colhead{}
}
\startdata
GSC 03122-02426&W&0.30&1.30&0.83&2.19&0.81&0.058&97.046&44\\
DN Cam&W&0.50&1.78&1.22&1.85&0.82&0.056&101.841&1,41\\
ER Ori&W&0.42&1.39&1.14&1.53&0.98&0.041&95.960&5,35\\
EF Boo&W&0.43&1.46&1.09&1.61&0.82&0.047&86.367&35,41\\
AA UMa&W&0.47&1.47&1.11&1.56&0.85&0.042&89.033&35,41\\
BI Cvn&W&0.38&1.37&0.92&1.59&0.65&0.054&67.387&28\\
RR Cen&W&0.61&2.10&1.05&1.82&0.38&0.105&55.342&37\\
YY Eri&W&0.32&1.20&0.77&1.55&0.62&0.054&59.627&7,41\\
FZ Ori&W&0.39&1.16&1.082&1.17&1.00&0.036&76.048&23\\
ET Leo&W&0.35&1.36&0.84&1.59&0.54&0.069&55.806&4,41\\
V502 Oph&W&0.45&1.51&0.93&1.51&0.56&0.058&59.990&7,41\\
W UMa&W&0.33&1.17&0.85&1.35&0.69&0.0467&59.373&7\\
AE Phe&W&0.36&1.29&0.81&1.38&0.63&0.052&57.474&7\\
UX Eri&W&0.45&1.45&0.91&1.45&0.54&0.056&55.873&27\\
V402 Aur&W&0.60&1.98&0.92&1.64&0.33&0.103&44.808&35,46\\
TY Pup&W&0.82&2.64&1.37&1.65&0.30&0.133&47.196&31\\
AW Vir&W&0.35&1.08&0.95&1.11&0.84&0.037&60.954&26\\
V728 Her&W&0.47&1.81&0.92&1.65&0.3&0.132&38.843&35\\
BB Peg&W&0.36&1.29&0.83&1.42&0.53&0.060&50.613&11\\
V417 Aql&W&0.37&1.31&0.84&1.40&0.50&0.062&47.971&35,41\\
AM Leo&W&0.37&1.23&0.85&1.29&0.59&0.045&51.551&41,47\\
VY Sex&W&0.44&1.50&0.86&1.42&0.45&0.068&47.013&4,41\\
UV Lyn&W&0.42&1.39&0.89&1.36&0.5&0.060&48.656&34,35\\
V781 Tau&W&0.41&1.21&0.85&1.29&0.57&0.043&51.488&10,36\\
IK Boo&W&0.30&0.91&0.85&0.99&0.86&0.035&53.483&14\\
CW Cas&W&0.32&1.09&0.75&1.25&0.56&0.049&45.809&9\\
RZ Com&W&0.34&1.12&0.78&1.23&0.55&0.048&45.413&7,41\\
AH Vir&W&0.41&1.41&0.84&1.36&0.41&0.073&40.798&35\\
SW Lac&W&0.32&1.03&0.94&0.98&0.78&0.042&50.284&35,41\\
QW Gem&W&0.36&1.26&0.75&1.31&0.44&0.065&40.256&15,35\\
EM Lac&W&0.39&1.19&0.97&1.06&0.67&0.043&50.143&26\\
EZ Hya&W&0.45&1.54&0.85&1.37&0.35&0.086&37.070&35\\
V842 Her&W&0.42&1.46&0.81&1.36&0.35&0.085&35.970&35\\
SS Ari&W&0.41&1.36&0.80&1.31&0.40&0.069&38.605&35\\
GZ And&W&0.31&1.01&0.74&1.12&0.59&0.044&43.216&1,35\\
V752 Cen&W&0.37&1.28&0.75&1.3&0.4&0.069&37.217&2,35\\
VY Cet&W&0.34&1.01&0.83&1.02&0.68&0.037&46.856&26\\
V1139 Cas&W&0.30&0.94&0.76&1.02&0.66&0.039&43.522&17\\
TW Cet&W&0.32&0.99&0.76&1.06&0.61&0.040&43.019&35,41\\
V1128 Tau&W&0.31&1.01&0.76&1.09&0.58&0.045&41.729&43\\
V870 Ara&W&0.40&1.67&0.61&1.5&0.12&0.317&16.553&32,41\\
AO Cam&W&0.33&1.09&0.73&1.12&0.49&0.050&37.837&1,3\\
CK Boo&W&0.36&1.45&0.59&1.39&0.16&0.215&17.710&40\\
U Peg&W&0.38&1.22&0.74&1.15&0.38&0.063&32.372&25,35\\
EP And&W&0.40&1.27&0.82&1.10&0.41&0.059&34.267&19\\
AB And&W&0.33&1.05&0.76&1.01&0.49&0.048&34.909&8\\
V1007 Cas&W&0.33&1.16&0.69&1.14&0.34&0.073&28.141&18\\
BV Dra&W&0.35&1.12&0.76&1.04&0.43&0.054&32.526&12,35\\
V757 Cen&W&0.34&0.97&0.80&0.88&0.59&0.037&36.905&35\\
FU Dra&W&0.31&1.12&0.61&1.17&0.29&0.086&24.393&34,35\\
GW Cep&W&0.32&1.05&0.67&1.06&0.39&0.057&29.363&26\\
GM Dra&W&0.34&1.25&0.61&1.21&0.22&0.120&20.532&3,41\\
V1191 Cyg&W&0.31&1.31&0.52&1.29&0.13&0.240&13.984&33\\
TX Cnc&W&0.39&1.28&0.91&1.32&0.60&0.051&54.093&20\\
BH Cas&W&0.41&0.85&0.96&0.45&0.47&0.049&33.296&This work\\
BX Peg&W&0.28&0.97&0.62&1.02&0.38&0.059&26.720&30,41\\
V789 Her&W&0.32&1.15&0.62&1.13&0.27&0.091&22.648&18\\
FU Dra&W&0.31&1.10&0.60&1.12&0.28&0.086&22.799&21\\
V546 And&W&0.38&1.23&0.66&1.08&0.28&0.079&23.469&6\\
UY UMa&W&0.38&1.40&0.63&1.19&0.16&0.173&16.242&42\\
TY Boo&W&0.32&1.00&0.69&0.93&0.40&0.052&27.006&29,41\\
VW Cep&W&0.28&0.91&0.62&0.93&0.40&0.051&25.827&7\\
AR Boo&W&0.34&1.00&0.65&0.90&0.35&0.050&23.972&16\\
V829 Her&W&0.36&1.07&0.74&0.86&0.37&0.053&24.723&35,46\\
XY Leo&W&0.28&0.83&0.66&0.76&0.46&0.040&24.900&32,41\\
OU Ser&W&0.30&1.09&0.51&1.02&0.18&0.124&14.429&24,35\\
V345 Gem&W&0.27&1.09&0.49&1.05&0.15&0.165&12.492&38\\
BW Dra&W&0.29&0.98&0.55&0.92&0.26&0.074&17.958&12,35\\
LR Cam&W&0.43&1.27&0.73&0.9&0.27&0.071&20.821&39\\
V523 Cas&W&0.23&0.74&0.55&0.75&0.38&0.045&19.616&43\\
CC Com&W&0.22&0.71&0.53&0.72&0.38&0.045&18.639&13\\
RW Dor&W&0.29&0.80&0.67&0.64&0.43&0.040&20.507&35\\
RW Com&W&0.24&0.71&0.46&0.56&0.20&0.061&8.971&22,35\\
\enddata
\end{deluxetable*}
\tablecomments{Ref.:\citet{2004AcA....54..195B},2:\citet{1993ApJ...407..237B},3:\citet{2005AcA....55..123G},4:\citet{2006AcA....56..127G},5:\citet{1994A&A...289..827G},
6:\citet{2015NewA...36..100G},7:\citet{1988MNRAS.231..341H},8:\citet{1988ApJ...335..319H},9:\citet{2010PASJ...62..457J},10:\citet{2006A&A...452..959K},11:\citet{2007AJ....134..642K},
12:\citet{1986AJ.....92..666K},13:\citet{2011AN....332..626K},14:\citet{2017PASJ...69...62K},15:\citet{2003A&A...412..465K},16:\citet{2009AJ....138..478L},17:\citet{2015NewA...34..217L},
18:\citet{2018PASP..130g4201L},19:\citet{2013AJ....146...79L},20:\citet{2007PASJ...59..607L},21:\citet{2012PASJ...64...48L},22:\citet{1987ApJ...319..325M},23:\citet{2014Ap&SS.353..575P},
24:\citet{1999A&A...345..137P},25:\citet{2002CoSka..32...79P},26:\citet{2003MNRAS.342.1260Q},27:\citet{2007AJ....134.1769Q},28:\citet{2008AJ....136.2493Q},29:\citet{1990MNRAS.246...42R},
30:\citet{1991AJ....102.1171S},31:\citet{2018AJ....156..199S},32:\citet{2007A&A...465..943S},33:\citet{2012NewA...17...46U},34:\citet{2001CoSka..31..129V},35:\citet{2005ApJ...629.1055Y},
36:\citet{2005MNRAS.363.1272Y},37:\citet{2005PASJ...57..983Y},38:\citet{2009AJ....138..540Y},39:\citet{2010PASJ...62.1045Y},40:\citet{2012AJ....143..122Y},41:\citet{2013MNRAS.430.2029Y},
42:\citet{2017NewA...55...13Y},43:\citet{2004MNRAS.347..307Z},44:\citet{2011RAA....11..583Z},45:\citet{2016NewA...48...12Z},46:\citet{2004AcA....54..299Z},47:\citet{2010MNRAS.408..464Z}}


\begin{thebibliography}{}



\bibitem[Andrae et al.(2018)]{2018A&A...616A...8A} Andrae, R., Fouesneau, M., Creevey, O., et al.\ 2018, \aap, 616, A8

\bibitem[Applegate(1992)]{1992ApJ...385..621A} Applegate, J.~H.\ 1992, \apj, 385, 621

\bibitem[Arbutina(2007)]{2007MNRAS.377.1635A} Arbutina, B.\ 2007, \mnras, 377, 1635
\bibitem[Arbutina(2009)]{2009MNRAS.394..501A} Arbutina, B.\ 2009, \mnras, 394, 501

\bibitem[Arranz Heras \& Sanchez-Bajo(2009)]{2009Obs...129...88A} Arranz Heras, T., \& Sanchez-Bajo, F.\ 2009, The Observatory, 129, 88


\bibitem[Baran et al.(2004)]{2004AcA....54..195B} Baran, A., Zola, S., Rucinski, S.~M., et al.\ 2004, \actaa, 54, 195


\bibitem[Binnendijk(1970)]{1970VA.....12..217B} Binnendijk, L.\ 1970, Vistas in Astronomy, 12, 217

\bibitem[Brandt et al.(1997)]{1997MNRAS.291..709B} Brandt, W.~N., Ward, M.~J., Fabian, A.~C., \& Hodge, P.~W.\ 1997, \mnras, 291, 709

\bibitem[Barone et al.(1993)]{1993ApJ...407..237B} Barone, F., di Fiore, L., Milano, L., \& Russo, G.\ 1993, \apj, 407, 237


\bibitem[Beljawsky(1931)]{1931AN....243..115B} Beljawsky, S.\ 1931, Astronomische Nachrichten, 243, 115

\bibitem[Chabrier et al.(2000)]{2000ApJ...542..464C} Chabrier, G., Baraffe, I., Allard, F., \& Hauschildt, P.\ 2000, \apj, 542, 464

\bibitem[Chen et al.(2006)]{2006AJ....131..990C} Chen, W.~P., Sanchawala, K., \& Chiu, M.~C.\ 2006, \aj, 131, 990

\bibitem[Covey et al.(2007)]{2007AJ....134.2398C} Covey, K.~R., Ivezi{\'c}, {\v Z}., Schlegel, D., et al.\ 2007, \aj, 134, 2398
\bibitem[Cox(2000)]{2000asqu.book.....C} Cox, A.~N.\ 2000, Allen's Astrophysical Quantities,
\bibitem[Fan et al.(2016)]{2016PASP..128k5005F} Fan, Z., Wang, H., Jiang, X., et al.\ 2016, \pasp, 128, 115005
\bibitem[Fox-Machado et al.(2018)]{2018PASP..130j4201F} Fox-Machado, L., Cang, T.~Q., Michel, R., Fu, J.~N., \& Li, C.~Q.\ 2018, \pasp, 130, 104201

\bibitem[Gaia Collaboration(2018)]{2018yCat.1345....0G} Gaia Collaboration 2018, VizieR Online Data Catalog, 1345,
\bibitem[Gazeas et al.(2005)]{2005AcA....55..123G} Gazeas, K.~D., Baran, A., Niarchos, P., et al.\ 2005, \actaa, 55, 123
\bibitem[Gazeas et al.(2006)]{2006AcA....56..127G} Gazeas, K.~D., Niarchos, P.~G., Zola, S., Kreiner, J.~M., \& Rucinski, S.~M.\ 2006, \actaa, 56, 127
\bibitem[Gazeas \& Niarchos(2006)]{2006MNRAS.370L..29G} Gazeas, K.~D., \& Niarchos, P.~G.\ 2006, \mnras, 370, L29

\bibitem[Goecking et al.(1994)]{1994A&A...289..827G} Goecking, K.-D., Duerbeck, H.~W., Plewa, T., et al.\ 1994, \aap, 289, 827

\bibitem[Gondoin(2004)]{2004A&A...415.1113G} Gondoin, P.\ 2004, \aap, 415, 1113

\bibitem[G{\"u}rol et al.(2015)]{2015NewA...36..100G} G{\"u}rol, B., Bradstreet, D.~H., \& Okan, A.\ 2015, \na, 36, 100


\bibitem[Harmanec(1988)]{1988BAICz..39..329H} Harmanec, P.\ 1988, Bulletin of the Astronomical Institutes of Czechoslovakia, 39, 329

\bibitem[Hilditch et al.(1988)]{1988MNRAS.231..341H} Hilditch, R.~W., King, D.~J., \& McFarlane, T.~M.\ 1988, \mnras, 231, 341

\bibitem[Hoffman et al.(2006)]{2006AJ....132.2260H} Hoffman, D.~I., Harrison, T.~E., McNamara, B.~J., et al.\ 2006, \aj, 132, 2260
\bibitem[Hrivnak(1988)]{1988ApJ...335..319H} Hrivnak, B.~J.\ 1988, \apj, 335, 319

\bibitem[Hu et al.(2016)]{2016AJ....151..170H} Hu, C.-P., Yang, T.-C., Chou, Y., et al.\ 2016, \aj, 151, 170

\bibitem[Huenemoerder et al.(2006)]{2006ApJ...650.1119H} Huenemoerder, D.~P., Testa, P., \& Buzasi, D.~L.\ 2006, \apj, 650, 1119
\bibitem[Huruhata(1952)]{1952PASP...64..200H} Huruhata, M.\ 1952, \pasp, 64, 200

\bibitem[Jansen et al.(2001)]{2001A&A...365L...1J} Jansen, F., Lumb, D., Altieri, B., et al.\ 2001, \aap, 365, L1
\bibitem[Jiang et al.(2010)]{2010PASJ...62..457J} Jiang, T.-Y., Li, L.-F., Han, Z.-W., \& Jiang, D.-K.\ 2010, \pasj, 62, 457
\bibitem[Kalberla et al.(2005)]{2005A&A...440..775K} Kalberla, P.~M.~W., Burton, W.~B., Hartmann, D., et al.\ 2005, \aap, 440, 775
\bibitem[Kallrath et al.(2006)]{2006A&A...452..959K} Kallrath, J., Milone, E.~F., Breinhorst, R.~A., et al.\ 2006, \aap, 452, 959
\bibitem[Kalomeni et al.(2007)]{2007AJ....134..642K} Kalomeni, B., Yakut, K., Keskin, V., et al.\ 2007, \aj, 134, 642
\bibitem[Kaluzny \& Rucinski(1986)]{1986AJ.....92..666K} Kaluzny, J., \& Rucinski, S.~M.\ 1986, \aj, 92, 666

\bibitem[Kandulapati et al.(2015)]{2015MNRAS.446..510K} Kandulapati, S., Devarapalli, S.~P., \& Pasagada, V.~R.\ 2015, \mnras, 446, 510

\bibitem[Kesseli et al.(2017)]{2017ApJS..230...16K} Kesseli, A.~Y., West, A.~A., Veyette, M., et al.\ 2017, \apjs, 230, 16



\bibitem[Kopal(1978)]{1978ASSL...68.....K} Kopal, Z.\ 1978, Astrophysics and Space Science Library, 68
\bibitem[K{\"o}se et al.(2011)]{2011AN....332..626K} K{\"o}se, O., Kalomeni, B., Keskin, V., Ula{\c s}, B., \& Yakut, K.\ 2011, Astronomische Nachrichten, 332, 626
\bibitem[Kriwattanawong et al.(2017)]{2017PASJ...69...62K} Kriwattanawong, W., Sanguansak, N., \& Maungkorn, S.\ 2017, \pasj, 69, 62
\bibitem[Kreiner et al.(2003)]{2003A&A...412..465K} Kreiner, J.~M., Rucinski, S.~M., Zola, S., et al.\ 2003, \aap, 412, 465

\bibitem[Kukarkin (1938)]{KukarkinB1938}Kukarkin,B. \ 1938, \ Veranderl.Sterne Nishni-Novgorod, 5,195

\bibitem[Lanza \& Rodon{\`o}(2002)]{2002AN....323..424L} Lanza, A.~F., \& Rodon{\`o}, M.\ 2002, Astronomische Nachrichten, 323, 424
\bibitem[Lee et al.(2008)]{2008AJ....135.1523L} Lee, J.~W., Youn, J.-H., Kim, C.-H., Lee, C.-U., \& Kim, H.-I.\ 2008, \aj, 135, 1523
\bibitem[Lee et al.(2009)]{2009AJ....138..478L} Lee, J.~W., Youn, J.-H., Lee, C.-U., Kim, S.-L., \& Koch, R.~H.\ 2009, \aj, 138, 478
\bibitem[Li et al.(2004)]{2004MNRAS.351..137L} Li, L., Han, Z., \& Zhang, F.\ 2004, \mnras, 351, 137
\bibitem[Li et al.(2015)]{2015NewA...34..217L} Li, K., Hu, S.-M., Guo, D.-F., et al.\ 2015, \na, 34, 217
\bibitem[Li et al.(2018)]{2018PASP..130g4201L} Li, K., Xia, Q.-Q., Hu, S.-M., Guo, D.-F., \& Chen, X.\ 2018, \pasp, 130, 074201
\bibitem[Li \& Zhang(2006)]{2006MNRAS.369.2001L} Li, L., \& Zhang, F.\ 2006, \mnras, 369, 2001
\bibitem[Liao et al.(2013)]{2013AJ....146...79L} Liao, W.-P., Qian, S.-B., Li, K., et al.\ 2013, \aj, 146, 79
\bibitem[Liu et al.(2007)]{2007PASJ...59..607L} Liu, L., Qian, S.-B., Boonrucksar, S., et al.\ 2007, \pasj, 59, 607
\bibitem[Liu et al.(2012)]{2012PASJ...64...48L} Liu, L., Qian, S.-B., He, J.-J., et al.\ 2012, \pasj, 64, 48

\bibitem[Lu \& Rucinski(1999)]{1999AJ....118..515L} Lu, W., \& Rucinski, S.~M.\ 1999, \aj, 118, 515


\bibitem[Lucy(1967)]{1967ZA.....65...89L} Lucy, L.~B.\ 1967, \zap, 65, 89
\bibitem[Ma et al.(2018)]{2018Ap&SS.363...68M} Ma, S.-G., Esamdin, A., Ma, L., et al.\ 2018, \apss, 363, 68

\bibitem[Mallik(1997)]{1997A&AS..124..359M} Mallik, S.~V.\ 1997, \aaps, 124, 359

\bibitem[McCartney(1997)]{1997ASPC..130..129M} McCartney, S.\ 1997, The Third Pacific Rim Conference on Recent Development on Binary Star Research, 130, 129
\bibitem[Milone et al.(1987)]{1987ApJ...319..325M} Milone, E.~F., Wilson, R.~E., \& Hrivnak, B.~J.\ 1987, \apj, 319, 325



\bibitem[Metcalfe(1999)]{1999AJ....117.2503M} Metcalfe, T.~S.\ 1999, \aj, 117, 2503

\bibitem[Milone(1968)]{1968AJ.....73..708M} Milone, E.~E.\ 1968, \aj, 73, 708

\bibitem[Mol{\'{\i}}k(1998)]{1998stel.conf...81M} Mol{\'{\i}}k, P.\ 1998, 20th Stellar Conference of the Czech and Slovak Astronomical Institutes, 81

\bibitem[M{\"u}ller(1903)]{1903PA.....11..138M} M{\"u}ller, G.\ 1903, Popular Astronomy, 11, 138


\bibitem[Niarchos et al.(2001)]{2001hell.confE..77N} Niarchos, P.~G., Aslanidis, D., Zola, S., \& Theodossiou, E.\ 2001, 5th Hellenic Astronomical Conference, 77.1

\bibitem[Nef \& Rucinski(2008)]{2008MNRAS.385.2239N} Nef, P.~D., \& Rucinski, S.~M.\ 2008, \mnras, 385, 2239

\bibitem[O'Connell(1951)]{1951PRCO....2...85O} O'Connell, D.~J.~K.\ 1951, Publications of the Riverview College Observatory, 2, 85

\bibitem[Pasham et al.(2013)]{2013ApJ...771L..44P} Pasham, D.~R., Strohmayer, T.~E., \& Mushotzky, R.~F.\ 2013, \apjl, 771, L44
\bibitem[Pi et al.(2017)]{2017AJ....154..260P} Pi, Q.-f., Zhang, L.-y., Bi, S.-l., et al.\ 2017, \aj, 154, 260

\bibitem[Prasad et al.(2014)]{2014Ap&SS.353..575P} Prasad, V., Pandey, J.~C., Patel, M.~K., \& Srivastava, D.~C.\ 2014, \apss, 353, 575
\bibitem[Pribulla et al.(1999)]{1999A&A...345..137P} Pribulla, T., Chochol, D., Rovithis-Livaniou, H., \& Rovithis, P.\ 1999, \aap, 345, 137
\bibitem[Pribulla \& Vanko(2002)]{2002CoSka..32...79P} Pribulla, T., \& Vanko, M.\ 2002, Contributions of the Astronomical Observatory Skalnate Pleso, 32, 79

\bibitem[Pr{\v s}a \& Zwitter(2005)]{2005ApJ...628..426P} Pr{\v s}a, A., \& Zwitter, T.\ 2005, \apj, 628, 426

\bibitem[Shengbang \& Qingyao(2000)]{2000Ap&SS.271..331S} Shengbang, Q., \& Qingyao, L.\ 2000, \apss, 271, 331

\bibitem[Qian(2001)]{2001MNRAS.328..635Q} Qian, S.\ 2001, \mnras, 328, 635
\bibitem[Qian(2003)]{2003MNRAS.342.1260Q} Qian, S.\ 2003, \mnras, 342, 1260
\bibitem[Qian et al.(2007)]{2007AJ....134.1769Q} Qian, S.-B., Yuan, J.-Z., Xiang, F.-Y., et al.\ 2007, \aj, 134, 1769
\bibitem[Qian et al.(2008)]{2008AJ....136.2493Q} Qian, S.-B., He, J.-J., Liu, L., Zhu, L.-Y., \& Liao, W.~P.\ 2008, \aj, 136, 2493

\bibitem[Qian et al.(2012)]{2012MNRAS.423.3646Q} Qian, S.-B., Zhang, J., Zhu, L.-Y., et al.\ 2012, \mnras, 423, 3646

\bibitem[Qian et al.(2014)]{2014ApJS..212....4Q} Qian, S.-B., Wang, J.-J., Zhu, L.-Y., et al.\ 2014, \apjs, 212, 4
\bibitem[Rainger et al.(1990)]{1990MNRAS.246...42R} Rainger, P.~P., Hilditch, R.~W., \& Bell, S.~A.\ 1990, \mnras, 246, 42

\bibitem[Riaz et al.(2018)]{2018MNRAS.478.5460R} Riaz, R., Vanaverbeke, S., \& Schleicher, D.~R.~G.\ 2018, \mnras, 478, 5460

\bibitem[Rovithis-Livaniou et al.(2000)]{2000A&A...354..904R} Rovithis-Livaniou, H., Kranidiotis, A.~N., Rovithis, P., \& Athanassiades, G.\ 2000, \aap, 354, 904

\bibitem[Ruci{\'n}ski(1969)]{1969AcA....19..245R} Ruci{\'n}ski, S.~M.\ 1969, \actaa, 19, 245


\bibitem[Samec \& Hube(1991)]{1991AJ....102.1171S} Samec, R.~G., \& Hube, D.~P.\ 1991, \aj, 102, 1171
\bibitem[Sarotsakulchai et al.(2018)]{2018AJ....156..199S} Sarotsakulchai, T., Qian, S.-B., Soonthornthum, B., et al.\ 2018, \aj, 156, 199


\bibitem[Shapley(1948)]{1948HarMo...7..249S} Shapley, H.\ 1948, Harvard Observatory Monographs, 7, 249
\bibitem[Song et al.(2016)]{2016RAA....16..154S} Song, F.-F., Esamdin, A., Ma, L., et al.\ 2016, Research in Astronomy and Astrophysics, 16, 154


\bibitem[St{\c e}pie{\'n} et al.(2001)]{2001A&A...370..157S} St{\c e}pie{\'n}, K., Schmitt, J.~H.~M.~M., \& Voges, W.\ 2001, \aap, 370, 157

\bibitem[Stone(1974)]{1974ApJ...193..135S} Stone, R.~P.~S.\ 1974, \apj, 193, 135

\bibitem[Str{\"u}der et al.(2001)]{2001A&A...365L..18S} Str{\"u}der, L., Briel, U., Dennerl, K., et al.\ 2001, \aap, 365, L18

\bibitem[Szalai et al.(2007)]{2007A&A...465..943S} Szalai, T., Kiss, L.~L., M{\'e}sz{\'a}ros, S., Vink{\'o}, J., \& Csizmadia, S.\ 2007, \aap, 465, 943

\bibitem[Tout \& Hall(1991)]{1991MNRAS.253....9T} Tout, C.~A., \& Hall, D.~S.\ 1991, \mnras, 253, 9
\bibitem[Turner et al.(2001)]{2001A&A...365L..27T} Turner, M.~J.~L., Abbey, A., Arnaud, M., et al.\ 2001, \aap, 365, L27
\bibitem[Ula{\c s} et al.(2012)]{2012NewA...17...46U} Ula{\c s}, B., Kalomeni, B., Keskin, V., K{\"o}se, O., \& Yakut, K.\ 2012, \na, 17, 46

\bibitem[van Hamme(1993)]{1993AJ....106.2096V} van Hamme, W.\ 1993, \aj, 106, 2096
\bibitem[Vanko et al.(2001)]{2001CoSka..31..129V} Vanko, M., Pribulla, T., Chochol, D., et al.\ 2001, Contributions of the Astronomical Observatory Skalnate Pleso, 31, 129

\bibitem[Vilhu \& Maceroni(2007)]{2007IAUS..240..719V} Vilhu, O., \& Maceroni, C.\ 2007, Binary Stars as Critical Tools  Tests in Contemporary Astrophysics, 240, Harmanec
\bibitem[Wilson \& Devinney(1971)]{1971ApJ...166..605W} Wilson, R.~E., \& Devinney, E.~J.\ 1971, \apj, 166, 605

\bibitem[Wilson(1979)]{1979ApJ...234.1054W} Wilson, R.~E.\ 1979, \apj, 234, 1054

\bibitem[Wilson(2012)]{2012AJ....144...73W} Wilson, R.~E.\ 2012, \aj, 144, 73

\bibitem[Yakut \& Eggleton(2005)]{2005ApJ...629.1055Y} Yakut, K., \& Eggleton, P.~P.\ 2005, \apj, 629, 1055
\bibitem[Yakut et al.(2005)]{2005MNRAS.363.1272Y} Yakut, K., Ula{\c s}, B., Kalomeni, B., \& G{\"u}lmen, {\"O}.\ 2005, \mnras, 363, 1272
\bibitem[Yang et al.(2005)]{2005PASJ...57..983Y} Yang, Y.-G., Qian, S.-B., Zhu, L.-Y., He, J.-J., \& Yuan, J.-Z.\ 2005, \pasj, 57, 983
\bibitem[Yang et al.(2009)]{2009AJ....138..540Y} Yang, Y.-G., Qian, S.-B., Zhu, L.-Y., \& He, J.-J.\ 2009, \aj, 138, 540
\bibitem[Yang \& Dai(2010)]{2010PASJ...62.1045Y} Yang, Y.-G., \& Dai, H.-F.\ 2010, \pasj, 62, 1045
\bibitem[Yang et al.(2012)]{2012AJ....143..122Y} Yang, Y.-G., Qian, S.-B., \& Soonthornthum, B.\ 2012, \aj, 143, 122
\bibitem[Yildiz \& Do{\u g}an(2013)]{2013MNRAS.430.2029Y} Yildiz, M., \& Do{\u g}an, T.\ 2013, \mnras, 430, 2029
\bibitem[Yu et al.(2017)]{2017NewA...55...13Y} Yu, Y.-X., Zhang, X.-D., Hu, K., \& Xiang, F.-Y.\ 2017, \na, 55, 13

\bibitem[Zhang \& Zhang(2004)]{2004MNRAS.347..307Z} Zhang, X.~B., \& Zhang, R.~X.\ 2004, \mnras, 347, 307
\bibitem[Zhang et al.(2011)]{2011RAA....11..583Z} Zhang, X.-B., Ren, A.-B., Luo, C.-Q., \& Luo, Y.-P.\ 2011, Research in Astronomy and Astrophysics, 11, 583
\bibitem[Zhou et al.(2016)]{2016NewA...48...12Z} Zhou, X., Qian, S.-B., He, J.-J., Zhang, J., \& Zhang, B.\ 2016, \na, 48, 12

\bibitem[Zo{\l}a et al.(2001)]{2001A&A...374..164Z} Zo{\l}a, S., Niarchos, P., Manimanis, V., \& Dapergolas, A.\ 2001, \aap, 374, 164
\bibitem[Zola et al.(2004)]{2004AcA....54..299Z} Zola, S., Rucinski, S.~M., Baran, A., et al.\ 2004, \actaa, 54, 299
\bibitem[Zola et al.(2010)]{2010MNRAS.408..464Z} Zola, S., Gazeas, K., Kreiner, J.~M., et al.\ 2010, \mnras, 408, 464



\end{thebibliography}
\end{document}